\documentclass[a4paper,11pt]{article}
\pdfoutput=1 % if your are submitting a pdflatex (i.e. if you have
\usepackage{jcappub} % for details on the use of the package, please
\usepackage[usenames,dvipsnames,table]{xcolor}
\usepackage{multirow}
\usepackage{makecell}
\usepackage{graphicx}
\usepackage{amssymb,amsmath,mathrsfs}
\usepackage{lscape}
\usepackage{xcolor,colortbl}
\usepackage{xspace}
\definecolor{Gray}{gray}{0.85}
\newcolumntype{a}{>{\columncolor{Gray}}c}
\newcolumntype{b}{>{\columncolor{white}}c}
\usepackage{verbatim}

\newcommand{\Mpl}{M_\mathrm{Pl}}

\newcommand{\GeV}{\mathrm{GeV}}
\newcommand{\MeV}{\mathrm{MeV}}
\newcommand{\eV}{\mathrm{eV}}
\newcommand{\keV}{\mathrm{keV}}

\newcommand{\beq}{\begin{equation}}
\newcommand{\eeq}{\end{equation}}

\newcommand{\Neff}{N_\mathrm{eff}}

\newcommand{\ds}{\mathrm{ds}}

\newcommand{\Nfld}{N_\mathrm{fld}}
\newcommand{\Ntot}{N_\mathrm{tot}}
\newcommand{\ffs}{f_\mathrm{fs}}
\newcommand{\fld}{\mathrm{fld}}

\newcommand{\montepython}{\texttt{MontePython}\xspace}
\newcommand{\class}{\texttt{CLASS}\xspace}
\newcommand{\camb}{\texttt{CAMB}\xspace}
\newcommand{\hc}{\mathrm{h.c.}}

\newcommand\KICP{Kavli Institute for Cosmological Physics, University of Chicago, Chicago, IL 60637, USA}
\newcommand\FNAL{Fermi National Accelerator Laboratory, Batavia, IL 60510, USA }
\newcommand{\UMD}{Maryland Center for Fundamental Physics, Department of Physics, University of Maryland, College Park, MD 20742, USA}
\newcommand{\JHU}{Department of Physics and Astronomy, Johns Hopkins University, Baltimore, MD 21218, USA}
\title{Interacting radiation after Planck and its implications for the Hubble Tension}
\author[a,b]{Nikita Blinov,}
\affiliation[a]{\FNAL}
\affiliation[b]{\KICP}
\author[c,d]{Gustavo Marques-Tavares}
\affiliation[c]{\JHU}
\affiliation[d]{\UMD}
\emailAdd{nblinov@fnal.gov}
\emailAdd{gusmt@umd.edu}
\date{}
\abstract{
	Standard cosmology predicts that prior to matter-radiation equality about $41\%$ of the energy 
	density was in free-streaming neutrinos. In many beyond Standard Model scenarios, however, 
	the amount and free-streaming nature of this component is modified. 
    For example, this occurs in models with new neutrino self-interactions or an additional dark sector with interacting light particles.
    We consider several extensions of the standard cosmology that include a non-free-streaming radiation component 
	as motivated by such particle physics models and use the final Planck data release to constrain them.  
    This release contains significant improvements in the polarization likelihood which plays an important role in distinguishing free-streaming from interacting radiation species.
	Fixing the total amount of energy in radiation to match the expectation from standard 
    neutrino decoupling we find that the fraction of free-streaming radiation must be $\ffs > 0.8$ at 95\% CL 
    (combining temperature, polarization and baryon acoustic oscillation data).
    Allowing for arbitrary contributions of free-streaming and interacting radiation, the effective number of 
    new non-free-streaming degrees of freedom is constrained to 
    be $\Nfld < 0.6$ at 95\% CL. 
	Cosmologies with additional radiation are also known to ease the discrepancy between the local measurement and CMB inference 
	of the current expansion rate $H_0$. We show that including a non-free-streaming 
    radiation component allows for a larger amount of total energy density in radiation, leading to a mild improvement 
	of the fit to cosmological data compared to previously discussed models with only a free-streaming component.
}
\proceeding{FERMILAB-PUB-20-114-A-T}
\begin{document}
	
	\maketitle
	\section{Introduction}
    \label{sec:intro}
	
	Cosmological measurements have become an important probe of Physics Beyond the Standard Model (BSM). 
	Most recently the Planck mission has significantly improved the measurements of the Cosmic Microwave Background (CMB), enabling the precise determination of cosmological parameters and comparison of the standard $\Lambda$CDM cosmology to its extensions~\cite{Aghanim:2018eyx}. 
	The CMB is a particularly powerful probe of BSM scenarios with new light degrees of freedom that are very weakly coupled to the Standard Model (SM), which are challenging to probe directly with terrestrial experiments. 
	These particles, however, are a generic feature of many ultra-violet completions and extensions of the SM including 
    additional neutrinos~\cite{Abazajian:2001nj}, dark sectors (DS)~\cite{Vogel:2013raa,Berlin:2017ftj}, 
	string axions~\cite{Halverson:2018xge} and axion-like particles~\cite{Baumann:2016wac}, and Goldstone bosons~\cite{Weinberg:2013kea,Millea:2015qra}.
	Their cosmological imprints would be a first glimpse into BSM physics, and it therefore constitutes 
	an important science driver of the current and future CMB experiments~\cite{Abazajian:2016yjj,Ade:2018sbj}.
	
	New contributions to the energy density of the Universe during recombination from extra relativistic degrees of freedom 
	would lead to observable changes in the CMB spectrum. 
	This extra amount of energy density has been historically parametrized by the number of effective neutrinos, $\Neff$,
	with the assumption that it affects the CMB in the same way as neutrinos.
	There are two important effects on the CMB power-spectrum~\cite{Bashinsky:2003tk,Hou:2011ec} due to neutrinos. First, 
	they significantly contribute to the expansion rate before matter-radiation equality (MRE), changing the sound horizon 
	and diffusion scales - a purely background cosmology effect, which affects 
	crude features of the CMB power spectrum. Second, the supersonic 
	propagation of neutrino perturbations induces a phase shift on the sound waves in the plasma which 
    manifests as a phase shift in the CMB and baryon acoustic oscillation (BAO) peaks~\cite{Bashinsky:2003tk,Baumann:2015rya}.
	
	While $\Neff$ was an appropriate parametrization of any new relativistic degree of freedom for earlier CMB experiments, 
	Planck's measurements are sensitive enough to probe both their contribution to the expansion rate and 
	the evolution of perturbations~\cite{Friedland:2007vv,Follin:2015hya,Baumann:2015rya}. The latter is very sensitive to the particle nature of the 
	relativistic species. In particular, the phase shift is a direct consequence 
	of the free-streaming nature of neutrinos. In many BSM scenarios, however, the extra radiation interacts 
	with itself or with dark matter, preventing free-streaming and causing the radiation to behave as an ideal relativistic fluid. In these cases $\Neff$ is an 
	inadequate parametrization of the new species.
	
	In this paper we study the impact of Planck's final data release and other 
	cosmological data sets on models with both free-streaming and interacting 
	radiation species with their energy densities parametrized by $\Neff$ and $\Nfld$, respectively. 
	Throughout this work we pay special attention to the impact of the non-free-streaming species on the tensions in the measurements of the 
	local expansion rate $H_0$ and 
	and the amplitude of matter fluctuations $\sigma_8$. The discrepancy between local~\cite{Riess:2019cxk,Wong:2019kwg} and high-redshift inferences~\cite{Aghanim:2018eyx,Abbott:2017smn,Addison:2017fdm,Schoneberg:2019wmt} of the local 
	expansion rate has a significance of $\sim 4-5\sigma$ depending on the data sets used (although some local measurements are in better agreement with the CMB~\cite{Freedman:2019jwv}).
	It is well known that additional 
	relativistic degrees of freedom can alleviate this tension by reducing the scale of the sound horizon~\cite{Bernal:2016gxb,Aylor:2018drw,Knox:2019rjx}. 
	A multitude of models have been proposed that achieve this by introducing additional energy either in the form of $\Neff$~\cite{Riess:2016jrr}, neutrinos with sizable self-interactions~\cite{Kreisch:2019yzn,Forastieri:2019cuf} or interactions with dark matter~\cite{DiValentino:2017oaw,Ghosh:2019tab}, decaying dark matter~\cite{Poulin:2016nat} 
	or early dark energy~\cite{Poulin:2018cxd,Poulin:2018dzj,Agrawal:2019lmo,Smith:2019ihp}. 
	Models with non-free-streaming radiation species studied in this work can be considered 
	as a limit of the interacting neutrino models, where the self-interactions are sufficiently large 
	(i.e. faster than the Hubble rate) during the times relevant for the CMB. 
	The $\sigma_8$ tension is milder~\cite{Abbott:2017wau,Hildebrandt:2018yau,Aghanim:2018eyx,Abbott:2018wzc,Hikage:2018qbn}, at about $\sim 2\sigma$. 
	However models with additional energy at around MRE motivated by $H_0$ tend to increase $\sigma_8$ and aggravate the tension. 
	This effect is lessened in models with interacting radiation, enabling it to potentially address both anomalies.
	
	This paper is organized as follows. In Sec.~\ref{sec:models} we show how 
	scenarios with new free-streaming or interacting radiation can arise from well-motivated particle physics models. 
	There, we argue that wide variety of free-streaming and interacting species 
	occur naturally in models with non-Abelian hidden sectors, or late equilibration 
	of light particles.
	These and many other models can be mapped onto an effective fluid description as detailed in Sec.~\ref{sec:boltzmann}; the resulting Boltzmann equations 
	can be easily implemented in a solver such as \class~\cite{Blas:2011rf} or \camb~\cite{Lewis:1999bs}.
	We then extend previous analyses of Refs.~\cite{Baumann:2015rya,Brust:2017nmv} by including the latest data~\cite{Aghanim:2019ame} and performing extensive 
	comparisons against the base cosmology and models with only free-streaming species in Sec.~\ref{sec:constraints}. 
	The inclusion of the latest data set is essential due to the significant improvements in the treatment of polarization and its relevance for the distinction between interacting and free-streaming radiation, parametrized by $\Nfld$ and $\Neff$, respectively. 
	In addition to models with freely varying $\Neff$ and $\Nfld$, we consider subsets of this parameter space motivated by the 
	different scenarios discussed in Sec.~\ref{sec:models}. In particular, we study the scenario in which $\Neff+\Nfld = 3.046$ 
	to show that Planck disfavors even one of the neutrino flavors having significant interactions at the time of recombination. We also consider variations of 
	the primordial helium fraction, $Y_p$, which allows for much larger radiation 
	densities if one ignores constraints from direct measurements of $Y_p$. We summarize our findings and conclude in Sec.~\ref{sec:conclusion}.

	\section{Models of Dark Radiation}
	\label{sec:models}
	In this section we present simple models of dark radiation that lead to different predictions for the contributions to $\Neff$ and to its interacting counterpart $\Nfld$. These models are not meant as a complete set of possibilities, but as examples that motivate searching for a wide range of $(\Neff, \, \Nfld)$, which parametrize the cosmic energy density in relativistic degrees of freedom (in addition to the photon).\footnote{
		See also \cite{DiValentino:2017oaw,Ghosh:2019tab,Escudero:2019gvw,Kreisch:2019yzn,Forastieri:2019cuf} for other examples of models with an interacting radiation component.} Explicitly, these quantities are 
    normalized such that $\Neff \approx 3.046$ in $\Lambda$CDM well after neutrino-photon decoupling and electron-positron annihilation (we use 
    the result of Ref.~\cite{Mangano:2005cc} -- other recent calculations, e.g., Refs.~\cite{Grohs:2015tfy,deSalas:2016ztq}, give similar values) 
	so the total radiation energy density is
	\beq
	\rho_{\mathrm{rad}} = \rho_\gamma + \rho_\nu + \rho_\ds = \rho_\gamma\left[1 + \frac{7}{8}(\Neff + \Nfld) \left(\frac{4}{11}\right)^{4/3}\right],
	\label{eq:neff_nfld_def}
	\eeq
	where $\rho_\gamma$, $\rho_\nu$ and $\rho_\ds$ are the energy densities in photons, neutrinos and any additional 
	dark sector states.
	Thus, $\Neff$ and $\Nfld$ are completely degenerate at the level of the background cosmology. 
	Following Ref.~\cite{Brust:2017nmv} we will often reparametrize these densities in terms of 
	\begin{align}
      \Ntot & = \Neff + \Nfld, \\
		\ffs & = \Neff/\Ntot,
	\end{align}
	so $\Ntot$ is the total number of effective relativistic species, and $\ffs$ is the 
	fraction of those that are free-streaming. This parametrization reduces the degeneracy between $\Neff$ and $\Nfld$, enabling a more 
	efficient Monte Carlo exploration of the parameter space.
	As discussed in the introduction, the degeneracy is also broken by the different evolution of the perturbations in the two fluids due to the 
	presence or absence of self-interactions. Below we discuss how such interactions can naturally arise in 
	particle physics models.
	
	\subsection{Non-Abelian Dark Radiation}
	\label{sec:non_abelian_dr}
	
	A simple candidate for dark radiation are gauge bosons of a new non-Abelian group with a very small gauge coupling~\cite{Jeong:2013eza,Buen-Abad:2015ova,Lesgourgues:2015wza,Buen-Abad:2017gxg}. The smallness of the gauge coupling implies an exponentially small confinement scale, and therefore the relevant degrees of freedom for cosmology are the gauge bosons (dark gluons), which are effectively massless. The gauge structure imposes a minimum amount of self-interactions controlled by the gauge coupling $g_d$. Even a tiny gauge coupling is sufficient to make this a perfect fluid, given that the interaction rate is approximately
	\begin{equation}
		\Gamma_d \sim \alpha_d^2 T_d,
	\end{equation}
    where $\alpha_d = g_d^2/(4 \pi)$ and $T_d$ is the temperature of the dark fluid, which we take to be comparable to the SM temperature. If we require that this interaction is in equilibrium before the photon temperature reaches a keV (and so well before the modes probed by the CMB enter the horizon) we find that
	\begin{equation}
		\alpha_d \gtrsim 10^{-12} \, .
	\end{equation}
	This shows that even very small gauge couplings lead to sufficiently large self-interactions to ensure that the dark gluons locally thermalize and thus behave as an ideal fluid.
	
	The contribution of dark gluons to $\Nfld$ depends on the temperature of the fluid $T_d$, and their number $N_d$, 
    which is equal to the number of generators of the non-Abelian group (for example, for a dark $SU(N)$, $N_d = N^2-1$).
	The temperature of the dark sector depends on how it interacts with the Standard Model and on details of cosmology at very high temperatures. 
	If it was ever in thermal contact with the SM (and ignoring a possible late reheating of the SM), but decoupled at a temperature 
	$T_f$, the dark sector temperature follows from entropy conservation
	\begin{equation}
		\frac{T_d}{T_\gamma} = \left[\frac{g_{*S}(T_\gamma)}{g_{*S}(T_f)}\right]^{1/3} ,
		\label{eq:decoupled_ds_temp}
	\end{equation}
    where $g_{*S}$ is the effective number of (entropy) relativistic species including the neutrino contribution. 
	Using Eq.~\ref{eq:neff_nfld_def} this leads to
	\beq
	\Nfld = \frac{8}{7}\left(\frac{11}{4} \right)^{4/3}\left(\frac{T_d}{T_\gamma}\right)^4N_d \, .
    \label{eq:nfld_from_dr}
	\eeq
	Note that thermal equilibrium at early times implies a \emph{minimum} dark sector temperature
	\beq
	\frac{T_d}{T_\gamma} \gtrsim 0.33\;\Rightarrow\; \Nfld \gtrsim 0.054 N_d \, , 
	\eeq
	where we evaluated Eq.~\ref{eq:decoupled_ds_temp} at $T_f \gg 100\;\GeV$ and $T_\gamma \sim 1\;\eV$. 
	This shows that even the 
	simplest $SU(2)$ non-Abelian dark sector ($N_d = 3$) implies $\Nfld \gtrsim 0.16$ (in the absence of non-SM entropy injections 
	at temperatures below the weak scale).
	
	In the above model of a decoupled DS, $\Neff = 3.046$ and $\Nfld$ can take on any positive value. 
	Next, we consider a model where instead $\Ntot = 3.046$ is fixed, but $\ffs$ 
	is allowed to vary.
	
	%%%%%%%%%%%%%%%%%%%%%%%%%%%%
	\subsection{Late Equilibration of a Dark Sector}
	\label{sec:late_equilibration}
	
	A dark sector could come into equilibrium with the Standard Model neutrinos after Big Bang Nucleosynthesis (BBN) as considered in~\cite{Chacko:2003dt,Chacko:2004cz,Berlin:2017ftj}. 
	This possibility has a number of interesting features. In particular it predicts no deviations in $\Ntot$ during BBN, and, as we will discuss, 
	can decrease the neutrino contribution to $\Ntot$ at CMB, effectively replacing it with new BSM particles. We will refer to this scenario as ``late equilibration''.
	
	A simple model that realizes this scenario involves adding a new massive scalar $\phi$ that directly couples to neutrinos $\nu$ via
	\begin{equation}
		\mathcal{L} \supset \frac{1}{2} \lambda_\nu \phi \nu \nu + \hc,
		\label{eq:majoron_int}
	\end{equation}
	where $\lambda_\nu$ is a dimensionless coupling constant (neutrino flavor 
	indices are suppressed). 
	Such interactions are characteristic of Majoron models where $\phi$ is part of the sector that generates neutrino masses~\cite{Berlin:2018ztp}.
	The scalar will come into equilibrium with the neutrinos through inverse decays $ \nu \nu \leftrightarrow \phi$ at a temperature~\cite{Berlin:2017ftj,Berlin:2018ztp,Berlin:2019pbq}
	\begin{equation}
      T_\phi^{\mathrm{eq}} \sim \left(\frac{ \lambda_\nu^2 m_\phi^2 \Mpl}{8 \pi}\right)^{1/3} \, ,
        \label{eq:ds_equilibration_temp}
	\end{equation}
	where we have assumed $T_\phi > m_\phi$ (inverse decay also requires $m_\phi > 2m_{\nu}$, where $m_{\nu}$ is the 
    mass of the neutrino(s) interacting with $\phi$).
	Equilibration conserves energy, so as long as $m_\phi < T_\phi < \MeV$ (i.e. equilibration occurs after neutrino-photon decoupling) and the initial DS 
	temperature is lower than that of the SM, the total amount of energy in relativistic species remains approximately constant, i.e. $\Neff + \Nfld \approx 3$, 
	and the successful prediction 
	standard BBN remain~\cite{Berlin:2019pbq}. The forward and inverse $\phi$ decays that enforce chemical equilibrium 
	can also tightly couple the $\phi$ and neutrino fluids, preventing free-streaming. The kinematics of the 
    $1\leftrightarrow 2$ processes suppress the requisite momentum isotropization rate by an additional factor of $(m_\phi/T)^2$ 
    compared to the decay/production rate discussed above~\cite{Chacko:2003dt,Hannestad:2005ex}. As a result, 
    the temperature at which neutrino stops free-streaming, $T_\phi^{\mathrm{nfs}}$, is parametrically smaller than the equilibration temperature:
    \beq
    T_\phi^{\mathrm{nfs}} \sim \left[\left(T_\phi^{\mathrm{eq}}\right)^3 m_\phi^2\right]^{1/5}.
    \label{eq:ds_nfs_temp}
    \eeq
    If this kinetic equilibrium persists 
	through recombination (i.e. $m_\phi \lesssim 0.1\;\eV$), this minimal scenario predicts 
	\beq
	\Ntot = \Neff + \Nfld \approx 3 \, ,
	\eeq
	and the fraction of free-streaming radiation $\ffs$ 
	\beq
	\ffs = \Neff/\Ntot \leq 1
	\eeq
	depends on how many neutrino flavors $\phi$ interacts with (i.e. the flavor structure of Eq.~\ref{eq:majoron_int}). 
	The limits 
	$\ffs\rightarrow 0$ and $\ffs\rightarrow 1$ correspond to $\phi$ interacting with all or none of the neutrino flavors, respectively.\footnote{If the $\ffs\rightarrow 0$ regime is obtained via decays and inverse decays $\nu\nu\leftrightarrow \phi$, then we must demand $m_\phi \gtrsim 2 m_{\nu_3} > 0.1\;\eV$, so the treatment of $\phi$ as an additional radiation species may break down near recombination.}
	The $\ffs\rightarrow 0$ scenario has been extensively studied under different assumptions about the temperature 
	scaling of the reaction rate responsible for the $\phi$-$\nu$ interactions~\cite{Cyr-Racine:2013jua,Archidiacono:2013dua,Oldengott:2014qra,Forastieri:2015paa,Oldengott:2017fhy,Lancaster:2017ksf,Forastieri:2019cuf,Escudero:2019gvw}. 
    The $\Nfld$ model simply corresponds to the limit in which the $\phi$-$\nu$ (or the $\nu$-$\nu$ scattering implied by Eq.~\ref{eq:majoron_int}) interactions are faster than Hubble for all scales probed by the observed CMB. Interestingly, couplings $\lambda_\phi$ and masses $m_\phi$ 
	implied by late equilibration 
	are small enough to (mostly) avoid stringent cosmological and laboratory bounds that constrain new neutrino interactions~\cite{Blinov:2019gcj}. 
	
	The scalar $\phi$ can also be a portal to a richer DS if $\phi$ couples to additional states more strongly than to neutrinos. 
	Then $\phi$-$\nu$ equilibration also brings those extra states into equilibrium, sharing the total 
	energy density between the neutrinos and the DS proportionally to the number of degrees of freedom in each sector. 
	If the number of DS states is much larger than 3, most of the energy will be in these new particles rather than neutrinos. 
	This is a realization of the ``neutrinoless universe''~\cite{Beacom:2004yd}, where the non-photon energy density is almost entirely in 
	non-free-streaming species, i.e. $\Ntot \approx \Nfld$ and $\ffs \approx 0$. If some of DS states are massive and 
	at some point become non-relativistic, their entropy is shared among the remaining states, increasing 
	their temperature relative to photons and thereby increasing $\Ntot$.
	
	A concrete realization of the above ideas can be constructed by assuming that $\phi$ couples to $N$ light Weyl fermions $\psi_i$:
	\begin{equation}
      \mathcal{L} \supset \frac{1}{2} \phi \sum_{i=1}^{3} \lambda_{\nu_i} \nu_i \nu_i + \frac{1}{2}\lambda_\psi \phi \sum_{i=1}^{N} \psi_i \psi_i +\hc\, ,
	\end{equation}
    where we took the $\phi$ coupling to dark radiation to be the same for all species, but allowed for mass eigenstate-dependence of $\lambda_\nu$. 
    Assuming $\lambda_\psi \gg \lambda_\nu$, as soon as $\phi$ enters thermal and chemical equilibrium with neutrinos through inverse decay $\nu \nu \leftrightarrow \phi$, the dark radiation will also enter equilibrium with the neutrinos via decays of $\phi \leftrightarrow \psi \psi$. 
    After equilibration, the temperature of the combined $\nu$+DS bath is determined from energy conservation~\cite{Berlin:2018ztp,Berlin:2019pbq}:
	\beq
  \frac{T_{\nu+\ds}}{T_\gamma}  = \left(\frac{4}{11}\right)^{1/3} \left(\frac{N_\nu}{N_\nu+N^\prime}\right)^{1/4},
	\label{eq:nu_ds_temp_after_eq}
	\eeq
    where $N_\nu$ is the number of neutrino mass eigenstates that interact with $\phi$ (corresponding to the number of non-negligible couplings $\lambda_{\nu_i}$) and 
    $N' = N + (4/7)$ counts the contributions of $\psi_i$ and $\phi$; we also assumed that before equilibration the DS temperature is $T_\ds/T_\gamma\lesssim 0.5$.
	In the limit $N^\prime \rightarrow 0$ we recover the standard result for neutrino temperature well after $e^{\pm}$ annihilation. 
	If $N'\gg N_\nu$, however, we see that $\rho_\nu/(\rho_\phi + \rho_\psi) = N_\nu/N'$ and the interacting neutrinos contribute negligibly compared to the 
	DS states. One can check using the definition of Eq.~\ref{eq:neff_nfld_def} that the temperature in Eq.~\ref{eq:nu_ds_temp_after_eq} guarantees that 
	$\Ntot \approx 3$. This continues to hold as long as all particles in equilibrium are relativistic. 
	If a number $N_h$ of ``heavy'' $\psi$ states become non-relativistic and decay or annihilate into the other lighter states, entropy conservation leads to an 
	increase of $T_{\nu+\ds}/T_\gamma$ and a corresponding increase in $\Ntot$~\cite{Berlin:2018ztp,Berlin:2019pbq}:
	\beq
	\Ntot \approx 3-N_\nu + N_\nu \left(\frac{N_\nu + N'}{N_\nu + N' - N_h}\right)^{1/3}.
    \label{eq:ntot_from_late_equilibration}
	\eeq
	Thus, multiple ``freeze-out'' events can further increase the total radiation density compared to photons.
    Depending on $N$, couplings and masses of the DS states, this scenario can realize a wide range of $\Neff$ and $\Nfld$, or just the subspace with $\Ntot \approx 3$. Note that in this scenario, even if $m_\phi > \keV$, and therefore $\phi$ is not directly relevant for the physics of the CMB, the dark radiation can still be sufficiently self-interacting through off-shell $\phi$ exchanges to be treated as a perfect fluid during the time relevant for CMB measurements (because $\lambda_\psi$ can be much larger than $\lambda_\nu$), while the neutrinos would be free-streaming (because the $\nu$-$\nu$ scattering rate through off-shell $\phi$ would be too small). 
    The alternative possibility of dark sector 
    self-interactions falling out of equilibrium during times relevant for the CMB has been considered in Ref.~\cite{Choi:2018gho}.

	%%%%%%%%%%%%%%%%%%%%%%%%%%%%%%%%%%%%%%%%%%%%%%%%%%%%%%%%
	
	\section{Impact of Interacting Radiation}
	\label{sec:boltzmann}
	We work with the perturbed Friedmann-Robertson-Walker spacetime~\cite{Ma:1995ey,Hu:2004xd} 
	\beq
	ds^2 = a^2(\eta) \left[ -(1+2A) d\eta^2 + (\delta_{ij} + 2h_{ij}) dx^i dx^j\right]
	\eeq
	where $\eta$ is the conformal time, and $A = \Psi$ and $h_{ij} = -\Phi$ in conformal Newtonian gauge ($A=0$ and $h_{ij} = h \hat{k}_i \hat{k}_j/2 + 3\eta_L(\hat{k}_i \hat{k}_j - \delta_{ij}/3)$ in synchronous gauge)
	are the metric perturbations in Fourier space. 
	We follow the notation of Ref.~\cite{Poulin:2016nat} below, which closely resembles that of the Boltzmann solver~\class~\cite{Blas:2011rf}.~\footnote{
    We used $\eta_L$ for the synchronous gauge metric perturbation instead of $\eta$ used in Refs.~\cite{Ma:1995ey,Hu:2004xd} to distinguish it from the conformal time.}
	
	\subsection{Boltzmann Equations}
	The evolution equations for the interacting radiation fluid follow from 
	conservation of the energy-momentum tensor~\cite{Ma:1995ey}.
	The background equation is 
	\beq
	\rho' + 4\left(\frac{a'}{a}\right) \rho = 0,
	\eeq
	where primes denote derivatives with respect to the conformal time $\eta$; 
	the solution is $\rho(\eta) = \rho_i a(\eta)^{-4}$, the same as for a free-streaming species. 
    We define $\rho_i$ in terms of $\Omega_\fld = \rho_i/\rho_{\mathrm{crit}}$, 
	the fluid density today normalized to the critical density, which, in turn, is defined by $\Nfld$:
	\beq
	\Omega_\fld = \frac{7}{8} \Nfld \left(\frac{4}{11}\right)^{4/3} \Omega_\gamma.
	\eeq
	
	The perturbation equations in conformal time are
	\begin{align}
		\delta' & = -\frac{4}{3}(\theta + \mathbf{m}_{\mathrm{cont}}) \label{eq:pert_evol_delta}\\
		\theta' & = \frac{1}{4}k^2 \delta + \mathbf{m}_{\mathrm{Euler}} \label{eq:pert_evol_theta}
	\end{align}
	where $\delta$ and $\theta$ are the density and velocity perturbations (see, e.g., Ref.~\cite{Ma:1995ey}) and $\mathbf{m}$ are gauge-dependent quantities appearing in continuity and Euler equations; their 
	values in the Newtonian and synchronous gauges are
	\begin{center}
		\begin{tabular}{|c|c|c|}
			\hline
			& Newtonian & Synchronous \\ 
			\hline
			$\mathbf{m}_{\mathrm{cont}}$ & $ -3 \Phi' $ & $h'/2$ \\
			$\mathbf{m}_{\mathrm{Euler}}$ & $k^2 \Psi$ & $0$\\
			\hline
		\end{tabular} 
	\end{center}
	Combining Eqs.~\ref{eq:pert_evol_delta} and~\ref{eq:pert_evol_theta} gives
	\beq
	\delta'' + \frac{1}{3} k^2 \delta = -\frac{4}{3}(\mathbf{m}_{\mathrm{Euler}} + \mathbf{m}_{\mathrm{cont}}'),
	\label{eq:second_order_eom}
	\eeq
	which makes it clear that disturbances in this fluid propagate with the sound speed $1/\sqrt{3}$ 
    (the equivalent expression for free-streaming species depends on shear and higher moments of the phase space 
    distribution preventing the same interpretation of the $k^2\delta/3$ term).

	The presence of non-free-streaming radiation also changes the initial conditions for the adiabatic perturbations. In particular in conformal Newtonian gauge the initial condition for the gravitational potential is~\cite{Ma:1995ey,Bashinsky:2003tk}
	\begin{equation}
		\Psi(\eta_i) = -\frac{2\zeta}{3(1+4R_\nu/15)},
		\label{eq:psi_ic}
	\end{equation}
    where $\zeta$ is the comoving curvature perturbation fixed by inflation, $R_\nu = \rho_\mathrm{fs} / \rho_{\mathrm{tot}}$ is the ratio of energy density in free-streaming species to the total energy density. From that we see that introducing an interacting radiation component lowers $R_\nu$, and thus increases the amplitude of the potential (for fixed $\zeta$).

	\subsection{Impact on the CMB Power Spectrum}
	The impact of free-streaming and interacting radiation has been extensively studied~\cite{Bond:1983hb,Bashinsky:2003tk,Hou:2011ec,Baumann:2015rya}. 
	Here we summarize the main physical quantities highlighted in these works which will help us to interpret the results of the 
	following sections.
	
	At the background level, free-streaming and interacting radiation 
	contribute to the expansion rate of the universe, affecting the physical 
	distances that shape the power spectrum. In particular, 
    key roles are played by the sound horizon $r_s$, the 
    photon diffusion scale $r_d$ and the angular diameter distance to last scattering $D_A$.
	
	The comoving sound horizon $r_s$ is
	\beq
	r_s = \int_0^{t_d} \frac{dt}{a(t)}c_s(a) = \int_0^{a_d} da \frac{c_s(a)}{a^2 H(a)},
	\label{eq:sound_horizon}
	\eeq
	where $c_s$ is the baryon-photon sound speed and $t_d$ ($a_d$) time (scale factor) at which the baryons decouple from the radiation.\footnote{We will not 
		distinguish between the end of the drag epoch and last scattering, which lead to sound horizons that are slightly different. Importantly, 
		their dependence on model parameters is identical. See Ref.~\cite{Knox:2019rjx} for a recent discussion.}
	This is an early-time quantity that depends on the contents of the universe prior to recombination through 
	the Hubble expansion rate $H(a)$. The sound horizon determines the spacing 
	of the peaks in the CMB power spectrum as we discuss below. 
	
	Temperature fluctuations in the baryon-photon plasma are exponentially damped at small scales due to photon diffusion 
	induced by their non-zero mean free path.
	The characteristic comoving damping scale $r_d$ is given by~\cite{Hu:1994uz,Zaldarriaga:1995gi,Hu:1996mn}
	\beq
	r_d^2 = \pi^2 \int_0^{a_d} \frac{da}{a^3 x_e n_e H \sigma_T} \left[\frac{R^2 + 16(1+R)/15}{6(1+R)^2}\right],
	\label{eq:damping_scale}
	\eeq
	where $x_e(a)$ is the free electron fraction, $\sigma_T$ the Thomson scattering cross-section and $R = 3\rho_b/(4\rho_\gamma)$.
    Modes that are physically smaller than this scale have exponentially suppressed power.
    This equation highlights two important points. First, the response of $r_d$ to additional contributions to the energy density during the 
  radiation era (entering through $H$) is weaker compared to $r_s$ (due to the square root in the definition of $r_d$)~\cite{Hou:2011ec,Aylor:2018drw,Knox:2019rjx}; this ensures 
    that $r_s/r_d$ changes if the early-time energy content is modified.
    Second, Eq.~\ref{eq:damping_scale} makes explicit a degeneracy between additional radiation and $x_e$. 
    In particular, close to recombination $x_e \propto (1-Y_p)$, so the effects of larger 
	energy densities on the damping scale can be partly compensated by varying $Y_p$~\cite{Hou:2011ec}. 
	In the standard cosmology $Y_p$ is not a free parameter, since it is completely determined 
	by the baryon and neutrino densities through standard Big Bang Nucleosynthesis. 
	
	Neither the sound horizon, nor the photon diffusion scale are directly observable. 
	Instead, observations constrain the projection of these length scales onto the sky which additionally depends on 
	the (comoving) angular diameter distance to the surface of last scattering, $D_A$:
	\beq
	D_A = \int_{a_d}^{1} \frac{da}{a^2 H(a)}. 
	\eeq
	The angular diameter distance is a late-time quantity, i.e., a function of the energy content 
	after recombination. However, the power spectrum is sensitive to the 
	scale of matter-radiation equality through the radiation driving effect, which 
	amplifies the modes that enter the horizon during radiation domination~\cite{Hu:2000ti}.
	This results in a large positive correlation between the matter density $\omega_m$ and the amount additional 
	radiation, as the data requires the redshift of matter-radiation equality to be approximately fixed~\cite{Hou:2011ec}.
    Since $D_A$ depends sensitively on $\omega_m$, additional radiation modifies this distance scale indirectly.

    The projected quantities imprinted onto the CMB power spectrum are $r_s/D_A$ and $r_d/D_A$. 
    As we describe below, $r_s/D_A$ determines the angular scale of the peaks in the CMB power spectrum, which are measured with 
    exquisite precision. This tight constraint ensures that a decrease of $r_s$ due to additional radiation must be 
    compensated by a decrease in $D_A$ which is achieved by increasing $H_0$. This is why additional radiation tends to 
    ameliorate the Hubble tension. However, since this extra energy density necessarily modifies $r_s/r_d$ as described above, 
    the CMB constraint on the angular diffusion scale $r_d/D_A$ prevents large departures from $\Ntot \approx 3$. 
    Note that the effects described so far stem from variations of the background cosmology which 
    do not distinguish free-streaming from interacting radiation.

	The difference between free-streaming and interacting radiation arises 
	at first order in perturbation theory. As discussed in Refs.~\cite{Bashinsky:2003tk,Baumann:2015rya}, the supersonic propagation of free-streaming radiation induces a phase shift into the CMB power spectrum. The peak positions are shifted by (in the flat sky approximation)
	\beq
	\ell_p \approx n (\pi -\delta\varphi) \frac{D_A}{r_s},
	\label{eq:phase-shift}
	\eeq
    where for modes with $k r_s\gg 1$ the phase shift $\delta \varphi$ is~\cite{Bashinsky:2003tk,Baumann:2015rya}
	\beq
	\delta\varphi \approx 0.191\pi \left(\frac{\rho_{\mathrm{fs}}}{\rho_{\mathrm{rad}}}\right),
	\eeq
	where $\rho_{\mathrm{fs}}$ is the energy density of free-streaming species. 
	Even if the free-streaming energy density is fixed, additional contributions to non-free-streaming components will 
	lower this phase shift.

Another effect from interacting radiation is to reduce the impact of the free-streaming component on the radiation driving envelope of the CMB. Modes that enter the horizon during radiation domination get a boost in their amplitude due to the rapidly decaying gravitational potential, leading to an increase in the oscillation amplitude proportional to the gravitational potential at horizon crossing~\cite{Hu:1994uz,Hu:1995en}. Free-streaming components reduce the size of the gravitational potential with respect to the primordial curvature perturbations (as seen from Eq.~\ref{eq:psi_ic}). Introducing an interacting component reduces $\ffs$, enhancing power at small scales (compared to an $\Neff$ model with the same $\Ntot$). This effect can partly compensate for the increased effect of diffusion damping, and therefore we expect slightly weaker CMB constraints on $\Nfld$ models compared to the ones with extra $\Neff$. 
	
	In the following section we discuss how the correlations implied by the above discussion 
	are realized in the Monte Carlo results, and their implications 
	for the cosmological tensions.

\section{Cosmological Constraints}
\label{sec:constraints}

In this section we study the cosmological constraints on three different extensions of $\Lambda$CDM:
\begin{itemize}
	\item $(\Ntot = 3.046,\, \ffs)$: We fix $\Ntot$ to the expected $\Lambda$CDM value $3.046$ and allow $\ffs$ to vary with flat prior in the range $0 \leq \ffs \leq 1$.
	\item $(\Neff = 3.046,\, \Nfld)$: We fix $\Neff$ to the expected $\Lambda$CDM value $3.046$ and allow $\Nfld$ to vary with flat prior in the range $ 0 \leq \Nfld \leq 2$.
    \item $(\Neff,\, \Nfld)$: We allow $\Ntot = \Neff + \Nfld$ and $\ffs = \Neff/\Ntot$ to vary independently with flat priors in the range $2 \leq \Ntot \leq 4.5$ and $0 \leq \ffs \leq 1$.
\end{itemize}
At the end of this section we will also explore the known degeneracy between $Y_p$, the 
primordial ${}^4$He fraction, and extra radiation~\cite{Hou:2011ec}, by 
allowing $Y_p$ to vary independently of its standard primordial nucleosynthesis prediction. 
We will present results that do not include any prior on $Y_p$ and evaluate the impact of imposing a prior from direct measurements.

In order to obtain constraints on the parameters of these models we included the equations~\ref{eq:pert_evol_delta},~\ref{eq:pert_evol_theta} for an interacting radiation species in the Boltzmann code~\class~\cite{Blas:2011rf} interfaced with \montepython 
3.2\footnote{We found that the initial implementation
of the 2018 Planck likelihoods did not 
include a Gaussian prior on a linear combination of 
parameters accounting for background from the kinetic and thermal Sunayev-Zeldovich effects~\cite{Aghanim:2019ame}. These backgrounds contribute 
to the high-$\ell$ part of the CMB power spectrum and are therefore somewhat 
degenerate with the impact of extra radiation on the damping tail. Including this prior had a mild ($\lesssim 0.5\sigma$) effect 
on parameter means, but a more significant effect on the best-fit values of $\chi^2$ as the nuisance parameters were driven to 
their boundaries without this prior. The Gaussian prior is included in our work and has been implemented in \montepython 3.3.}~\cite{Brinckmann:2018cvx} using combinations of the following likelihoods:
\begin{itemize}
	\item {\it Planck} TT: Full Planck 2018 TT likelihood, low-$\ell$ polarization likelihood (lowE) and lensing likelihood as described in Ref.~\cite{Aghanim:2019ame}.
	\item {\it Planck} TT, TE, EE: Full Planck 2018 TT, TE and EE likelihoods and lensing likelihood as described in Ref.~\cite{Aghanim:2019ame}.
	\item BAO: BOSS DR12 BAO results~\cite{Alam:2016hwk} (both the BAO and full-shape results); small $z$ BAO measurements from the 6dF~\cite{Beutler:2011hx} and BOSS MGS~\cite{Ross:2014qpa} catalogs.
    \item $H_0$: local $H_0$ measurement (with a simple Gaussian likelihood parametrized by $H_0 = 74.03\pm 1.42$ km/s/Mpc) from Ref.~\cite{Riess:2019cxk}.
\end{itemize}
In addition to the new parameters associated with each model, we use flat priors for the $\Lambda$CDM parameters $(\theta_{s}, \omega_{b}, \omega_{c}, \ln10^{10}A_{s}, n_{s}, \tau)$, 
where $\omega_i = \Omega_i h^2$. These parameters are described in, e.g., Ref.~\cite{Aghanim:2018eyx}. 
Here we use $\Lambda$CDM to refer to the standard cosmological model with \emph{massless} neutrinos to facilitate comparisons against models with arbitrary radiation content; 
non-zero neutrino masses tend to worsen the tension of CMB data with the local measurement of $H_0$~\cite{Aghanim:2018eyx}, and would make the preference for additional radiation even more dramatic.
The results are obtained by running 8 chains for each model 
until the Gelman-Rubin convergence criterion $R -1\ll 0.05$ is satisfied for all of the parameters~\cite{Gelman:1992zz,Brooks:1998ab}. 
We compare the goodness-of-fit of various models via $\chi^2$ evaluated at the maximum of the posterior distribution, which 
is found numerically using multiple restarts of \texttt{scipy}~\cite{2020SciPy-NMeth} Nelder-Mead and \texttt{iminuit}~\cite{iminuit,1975CoPhC..10..343J} minimization.
We present our main findings in the next subsections; additional results including different data sets and the full 2d posterior 
distributions are collected in Appendices~\ref{sec:tt_only_results} and~\ref{sec:posterior_distributions}. 

\subsection{Testing the free-streaming hypothesis for neutrinos}

\begin{figure}
	\centering
	\includegraphics[width=0.5\textwidth]{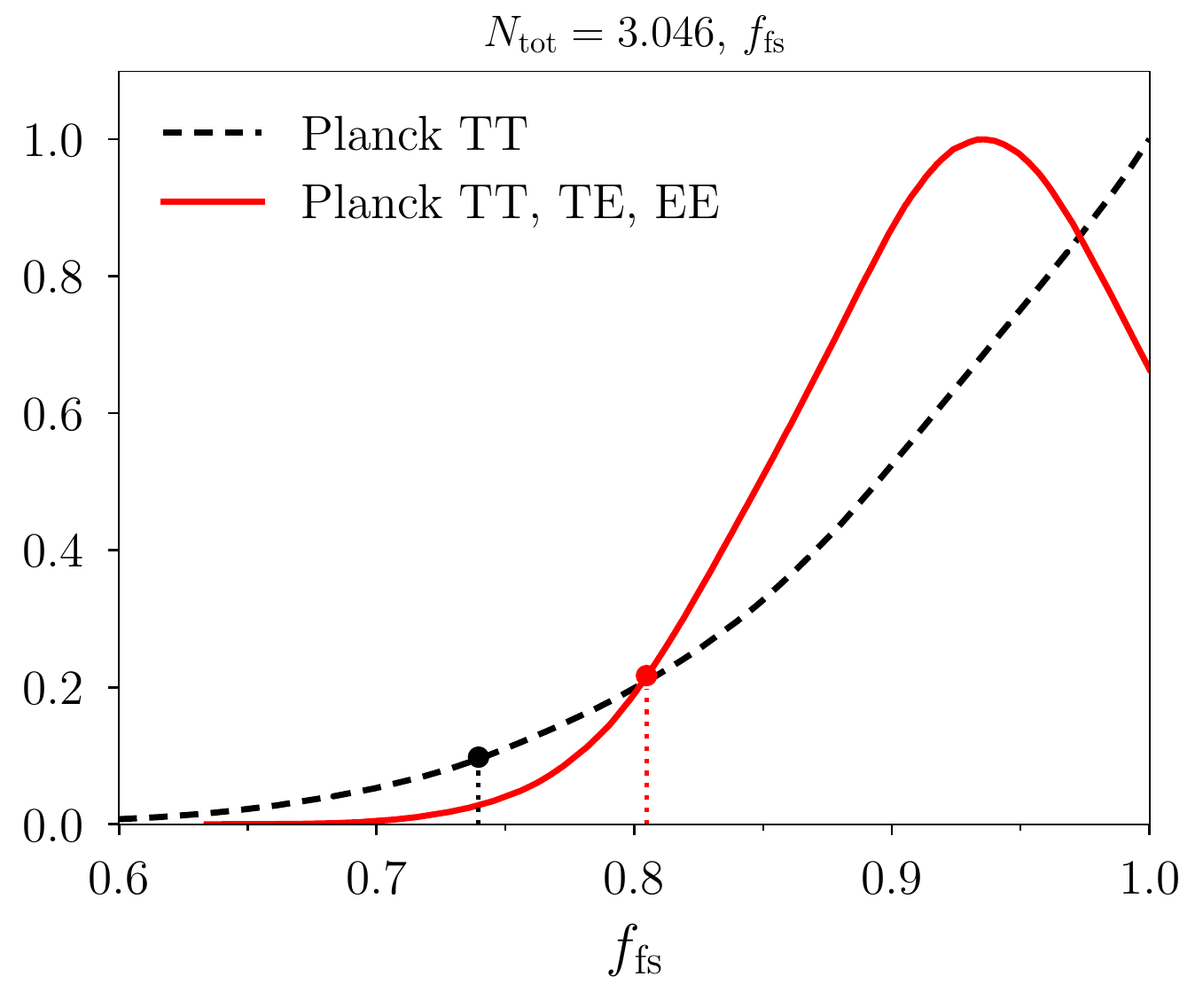}
    \caption{Marginalized posteriors of the fraction of energy in free-streaming radiation, $\ffs$, in the model with fixed $\Ntot=3.046$ using only 
      Planck data. The dashed black and solid red lines correspond to the results using {\it Planck} TT, and 
    TT, TE, EE data sets, respectively. The solid points denote the 95.4\% CL lower limit on $\ffs$ for each data set taken from Tab.~\ref{tab:ntot_fixed_results}.}
	\label{fig:ntot-fixed-ffs-1d}
\end{figure}

We first present results for the model with $\Ntot = 3.046$ with different
combinations of likelihoods in Table~\ref{tab:ntot_fixed_results}, where in
addition to the parameters scanned by \montepython we also show the derived
parameters $(H_0, r_s^{\mathrm{drag}},\sigma_8)$, where $r_s^{\mathrm{drag}}$ is the 
sound horizon at the end of the baryon drag epoch. We see that even using only the TT
likelihood there are already stringent constraints on the fraction of radiation
that is not free-streaming: $\ffs > 0.74$ (at $95.4 \%$ confidence). When we
include full polarization information and the BAO likelihood we find $\ffs >0.80$ (at $95.4 \%$ confidence), 
which shows that not even one of the neutrino
flavors can have significant interactions during recombination. This is in
agreement with a previous analysis of 2015 data by the Planck collaboration which
used a generalized fluid description to capture these different
possibilities~\cite{Ade:2015xua}. The posterior distribution for $\ffs$ is
shown in Fig.~\ref{fig:ntot-fixed-ffs-1d}, where we compare the effect of using
only the TT likelihood with using the full polarization data. In each
case there is no preference for $\ffs < 1$ at greater than $1\sigma$. Such a
preference emerges if one combines the CMB data with the local $H_0$
measurement; however, because this model does not have any additional energy
density and therefore does not address the $H_0$ tension, the two data sets are discrepant at $> 2\sigma$ and the small preference for non-zero interacting radiation component is irrelevant. 

\begin{table}
	\centering
	\begin{tabular}{|l|c|c|c|} 
\hline
 &  {\it Planck} TT  &  {\it Planck} TT, TE, EE  &  {\it Planck} TT, TE, EE + BAO \\
\hline
$100\theta_s$  &  $1.0427^{+0.0006}_{-0.00097}$  &  $1.0428^{+0.00054}_{-0.00073}$  &  $1.0428^{+0.00049}_{-0.00076}$\\
$100\Omega_b h^2$  &  $2.226^{+0.023}_{-0.025}$  &  $2.244^{+0.017}_{-0.015}$       &  $2.244^{+0.014}_{-0.014}$\\
$\Omega_c h^2$    &  $0.1207^{+0.0018}_{-0.0018}$  &  $0.1201^{+0.0012}_{-0.0012}$  &  $0.11996^{+0.00093}_{-0.00096}$\\
$\ln 10^{10}A_s$  &  $3.023^{+0.021}_{-0.017}$   &  $3.029^{+0.017}_{-0.018}$       &  $3.03^{+0.019}_{-0.016}$\\
$n_s$           &  $0.9561^{+0.009}_{-0.0064}$   &  $0.9589^{+0.0058}_{-0.0054}$    &  $0.9591^{+0.0062}_{-0.0051}$\\
$\tau$         &  $0.0525^{+0.0077}_{-0.0079}$   &  $0.0552^{+0.0073}_{-0.0077}$    &  $0.0551^{+0.0072}_{-0.0076}$\\
$\ffs$          &  $>0.74$     &  $>0.80$        &  $>0.80$\\
\hline
$H_0$ [km/s/Mpc]              &  $67.82^{+0.77}_{-0.79}$   &  $68.24^{+0.6}_{-0.57}$          &  $68.27^{+0.47}_{-0.41}$\\
$r_s^\mathrm{drag}$ [Mpc]         &  $147.04^{+0.53}_{-0.41}$  &  $147.02^{+0.27}_{-0.27}$  &  $147.03^{+0.24}_{-0.24}$\\
$\sigma_8$          &  $0.8168^{+0.0078}_{-0.0073}$  &  $0.8173^{+0.0072}_{-0.0073}$  &  $0.8171^{+0.0078}_{-0.0068}$\\
\hline
$\chi^2_{\mathrm{tot}}$ & $1189.06$ & $2774.07$ & $2780.92$ \\
\hline
	\end{tabular} 
    \caption{Results for a model with $\Ntot = 3.046$ with varying $\ffs$ fraction of free-streaming radiation species. 
      We present the marginalized mean $\pm 1 \, \sigma$ error for all the parameters, with the exception of $\ffs$ for which we show the $95\% $ lower limit. 
      The $\chi^2_{\mathrm{tot}}$ values correspond to best-fit point found using numerical minimization.
	\label{tab:ntot_fixed_results}}
\end{table}

\subsection{Models with varying \texorpdfstring{$\Ntot$}{Ntot}}

The results for models with varying $\Ntot$ are presented in
Tables~\ref{tab:ntot_planck} ({\it Planck} TT, TE, EE),~\ref{tab:ntot_planck_bao} ({\it Planck} TT, TE, EE + BAO)
and~\ref{tab:ntot_planck_bao_H0} ({\it Planck} TT, TE, EE + BAO + $H_0$), with $\Lambda$CDM and an $\Neff$ model
included for completeness (see Appendix~\ref{sec:tt_only_results} for results using only the TT likelihood). 
We highlight some of the interesting features below. 
\begin{itemize}
  \item CMB data requires the presence of free streaming species with at least $\ffs > 0.8-0.9$ depending on the model 
    and data set
  \item CMB and BAO data do not show preference for \emph{additional} interacting or free-streaming species
  \item When combined with the local measurement of $H_0$, a strong preference for additional radiation emerges
  \item This improvement comes at the price of a worse fit to the high-$\ell$ multipoles in part due to the larger effect of diffusion damping   
  \item Non-free-streaming allows for a slightly larger radiation component resulting in a mild improvement in fitting $H_0$ compared 
    to models with only free-streaming radiation.
\end{itemize}
We elaborate on these observations below.

\begin{table}
	\centering
	\begin{tabular}{|l|c|c|c|c|}
		\hline
		&  $\Lambda$CDM  &  $\Neff$  &  $\Neff=3.046$, $\Nfld$  &  $\Ntot$, $\ffs$ \\
		\hline
		$100\theta_s$  &  $1.0419^{+0.0003}_{-0.0003}$  &  $1.0422^{+0.00054}_{-0.00057}$  &  $1.042^{+0.00032}_{-0.00032}$  &  $1.0429^{+0.00062}_{-0.0007}$\\
		$100\Omega_b h^2$  &  $2.237^{+0.015}_{-0.015}$  &  $2.223^{+0.024}_{-0.021}$      &  $2.254^{+0.017}_{-0.021}$      &  $2.236^{+0.024}_{-0.025}$\\
		$\Omega_c h^2$    &  $0.1199^{+0.0011}_{-0.0013}$  &  $0.1176^{+0.0031}_{-0.003}$  &  $0.1221^{+0.0016}_{-0.0025}$   &  $0.1187^{+0.003}_{-0.0032}$\\
		$\ln 10^{10}A_s$  &  $3.043^{+0.014}_{-0.015}$  &  $3.036^{+0.017}_{-0.018}$       &  $3.044^{+0.014}_{-0.015}$      &  $3.026^{+0.019}_{-0.019}$\\
		$n_s$           &  $0.9649^{+0.0046}_{-0.0038}$  &  $0.9585^{+0.0091}_{-0.0089}$   &  $0.9673^{+0.0042}_{-0.0048}$   &  $0.9563^{+0.0084}_{-0.0083}$\\
		$\tau$         &  $0.054^{+0.0068}_{-0.0076}$   &  $0.0534^{+0.0071}_{-0.0079}$    &  $0.0557^{+0.0073}_{-0.0081}$   &  $0.0546^{+0.0069}_{-0.0082}$\\
		$\Ntot$         &  $3.046$                      &  $2.89^{+0.21}_{-0.2}$           &  $3.197^{+0.037}_{-0.151}$      &  $2.96^{+0.2}_{-0.21}$\\
		%$\ffs$          &  $1$                          &  $1$                             &  $0.953^{+0.045}_{-0.011}$      &  $0.917^{+0.083}_{-0.023}$\\
		$\ffs$          &  $1$                          &  $1$                             &  $> 0.89 $      &  $> 0.81$\\
		\hline
        $H_0$ [km/s/Mpc]               &  $67.93^{+0.54}_{-0.55}$  &  $66.8^{+1.6}_{-1.5}$            &  $69.14^{+0.77}_{-1.26}$        &  $67.7^{+1.4}_{-1.7}$\\
        $r_s^\mathrm{drag}$ [Mpc]         &  $147.14^{+0.27}_{-0.26}$  &  $148.8^{+1.9}_{-2.1}$  &  $145.6^{+1.5}_{-0.6}$         &  $147.9^{+2.0}_{-2.0}$\\
		$\sigma_8$          &  $0.8232^{+0.0057}_{-0.0059}$  &  $0.816^{+0.01}_{-0.011}$   &  $0.8264^{+0.0066}_{-0.0069}$   &  $0.814^{+0.01}_{-0.011}$\\
		\hline
		$\chi^2_{\mathrm{tot}}$ & 2774.75  &  2772.88  &  2774.75  &  2772.79 \\
		\hline
	\end{tabular}
	\caption{Comparison of extensions of $\Lambda$CDM with extra radiation degrees of freedom for the {\it Planck} TT, TE, EE likelihood.
    We present the marginalized mean $\pm 1 \, \sigma$ error for all the parameters, with the exception of $\ffs$ for which we show the $95\% $ lower limit. 
      The $\chi^2_{\mathrm{tot}}$ values correspond to best-fit point found using numerical minimization.
	\label{tab:ntot_planck}}
\end{table}

\begin{table}
	\centering
	\begin{tabular}{|l|c|c|c|c|}
		\hline
		&  $\Lambda$CDM  &  $\Neff$  &  $\Neff=3.046$, $\Nfld$  &  $\Ntot$, $\ffs$ \\
		\hline
		$100\theta_s$  &  $1.0419^{+0.00027}_{-0.00031}$  &  $1.0422^{+0.00053}_{-0.0005}$  &  $1.042^{+0.0003}_{-0.0003}$  &  $1.0429^{+0.00063}_{-0.00075}$\\
		$100\Omega_b h^2$  &  $2.239^{+0.014}_{-0.013}$   &  $2.232^{+0.017}_{-0.02}$       &  $2.252^{+0.015}_{-0.017}$    &  $2.24^{+0.019}_{-0.021}$\\
		$\Omega_c h^2$    &  $0.11964^{+0.0009}_{-0.00092}$  &  $0.118^{+0.0027}_{-0.0031}$  &  $0.122^{+0.0014}_{-0.0026}$  &  $0.1191^{+0.0031}_{-0.0032}$\\
		$\ln 10^{10}A_s$  &  $3.044^{+0.012}_{-0.016}$    &  $3.039^{+0.016}_{-0.015}$      &  $3.043^{+0.013}_{-0.014}$    &  $3.027^{+0.018}_{-0.018}$\\
		$n_s$           &  $0.9655^{+0.0037}_{-0.0038}$   &  $0.9621^{+0.007}_{-0.0072}$    &  $0.9671^{+0.0039}_{-0.0042}$  &  $0.9573^{+0.0081}_{-0.0077}$\\
		$\tau$         &  $0.0545^{+0.0065}_{-0.0076}$    &  $0.0543^{+0.0068}_{-0.0075}$   &  $0.0553^{+0.0068}_{-0.0072}$  &  $0.0549^{+0.0068}_{-0.0077}$\\
		$\Ntot$         &  $3.046$                        &  $2.94^{+0.16}_{-0.19}$         &  $3.192^{+0.032}_{-0.146}$    &  $2.99^{+0.18}_{-0.19}$\\
		%$\ffs$          &  $1$                            &  $1$                            &  $0.954^{+0.044}_{-0.009}$    &  $0.913^{+0.081}_{-0.027}$\\
		$\ffs$          &  $1$                            &  $1$                            &  $>0.89$    &  $> 0.80$\\
		\hline
        $H_0$ [km/s/Mpc]               &  $68.05^{+0.44}_{-0.41}$    &  $67.4^{+1.1}_{-1.2}$           &  $69.09^{+0.62}_{-1.08}$      &  $67.9^{+1.2}_{-1.3}$\\
        $r_s^\mathrm{drag}$ [Mpc]         &  $147.18^{+0.23}_{-0.21}$  &  $148.3^{+1.8}_{-1.7}$  &  $145.7^{+1.5}_{-0.6}$        &  $147.6^{+1.9}_{-1.9}$\\
		$\sigma_8$          &  $0.8227^{+0.0057}_{-0.0062}$  &  $0.8178^{+0.0094}_{-0.0102}$  &  $0.8261^{+0.0063}_{-0.0066}$  &  $0.815^{+0.01}_{-0.01}$\\
		\hline
		$\chi^2_{\mathrm{tot}}$ & 2781.06  &  2780.95  &  2781.04  &  2780.95 \\
		\hline
	\end{tabular}
	\caption{Comparison of extensions of $\Lambda$CDM with extra radiation degrees of freedom for {\it Planck} TT, TE, EE + BAO likelihood.
        We present the marginalized mean $\pm 1 \, \sigma$ error for all the parameters, with the exception of $\ffs$ for which we show the $95\% $ lower limit. 
      The $\chi^2_{\mathrm{tot}}$ values correspond to best-fit point found using numerical minimization. }
	\label{tab:ntot_planck_bao}
\end{table}

\begin{table}
	\centering
	\begin{tabular}{|l|c|c|c|c|}
		\hline
		&  $\Lambda$CDM  &  $\Neff$  &  $\Neff=3.046$, $\Nfld$  &  $\Ntot$, $\ffs$ \\
		\hline
		$100\theta_s$  &  $1.042^{+0.00029}_{-0.00028}$  &  $1.0414^{+0.00043}_{-0.00049}$  &  $1.0423^{+0.0003}_{-0.00032}$  &  $1.0427^{+0.00074}_{-0.00088}$\\
		$100\Omega_b h^2$  &  $2.249^{+0.013}_{-0.015}$  &  $2.265^{+0.017}_{-0.016}$       &  $2.275^{+0.016}_{-0.018}$      &  $2.275^{+0.018}_{-0.017}$\\
		$\Omega_c h^2$    &  $0.11861^{+0.00093}_{-0.00092}$  &  $0.1229^{+0.0026}_{-0.0026}$  &  $0.1248^{+0.0026}_{-0.0029}$  &  $0.1244^{+0.0029}_{-0.0029}$\\
		$\ln 10^{10}A_s$  &  $3.049^{+0.014}_{-0.015}$   &  $3.058^{+0.015}_{-0.016}$       &  $3.043^{+0.014}_{-0.016}$      &  $3.036^{+0.021}_{-0.019}$\\
		$n_s$           &  $0.9681^{+0.0035}_{-0.004}$   &  $0.9761^{+0.0061}_{-0.0057}$    &  $0.9704^{+0.004}_{-0.0038}$    &  $0.9669^{+0.0082}_{-0.0075}$\\
		$\tau$         &  $0.0576^{+0.007}_{-0.0078}$    &  $0.0575^{+0.007}_{-0.0076}$     &  $0.0574^{+0.0066}_{-0.0082}$   &  $0.0575^{+0.007}_{-0.0074}$\\
		$\Ntot$         &  $3.046$                       &  $3.3^{+0.15}_{-0.15}$           &  $3.38^{+0.13}_{-0.15}$         &  $3.35^{+0.16}_{-0.15}$\\
		$\ffs$          &  $1$                           &  $1$                             &  $0.901^{+0.039}_{-0.036}$      &  $0.87^{+0.08}_{-0.06}$\\
		\hline
        $H_0$ [km/s/Mpc]               &  $68.55^{+0.46}_{-0.41}$   &  $70.0^{+0.93}_{-0.9}$           &  $70.64^{+0.93}_{-1.0}$         &  $70.5^{+1.0}_{-1.0}$\\
        $r_s^\mathrm{drag}$ [Mpc]          &  $147.34^{+0.22}_{-0.23}$  &  $144.8^{+1.4}_{-1.5}$  &  $143.9^{+1.5}_{-1.3}$          &  $144.1^{+1.5}_{-1.6}$\\
		$\sigma_8$          &  $0.8216^{+0.0061}_{-0.0062}$  &  $0.8335^{+0.0089}_{-0.0093}$  &  $0.8299^{+0.0068}_{-0.0078}$  &  $0.826^{+0.011}_{-0.01}$\\
		\hline
		$\chi^2_{\mathrm{tot}}$ & 2797.46  &  2795.16  &  2793.65  &  2793.44 \\
		\hline
	\end{tabular}
	\caption{Comparison of extensions of $\Lambda$CDM with extra radiation degrees of freedom for the {\it Planck} TT, TE, EE + BAO + $H_0$ likelihood.
        We present the marginalized mean $\pm 1 \, \sigma$ error for all the parameters. 
      The $\chi^2_{\mathrm{tot}}$ values correspond to best-fit point found using numerical minimization.}
	\label{tab:ntot_planck_bao_H0}
\end{table}

\begin{figure}
	\centering
	\includegraphics[width=0.47\textwidth]{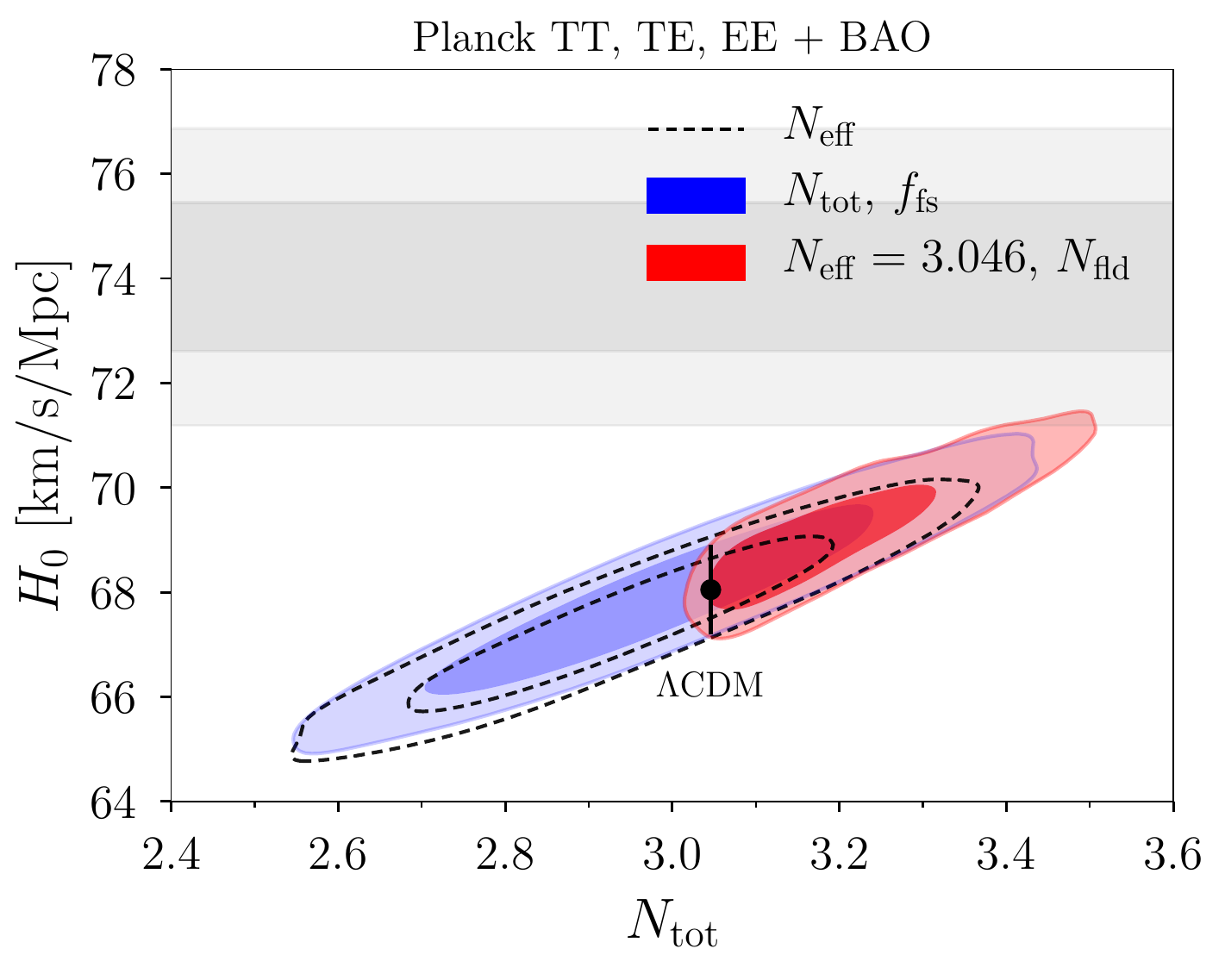}
	\includegraphics[width=0.47\textwidth]{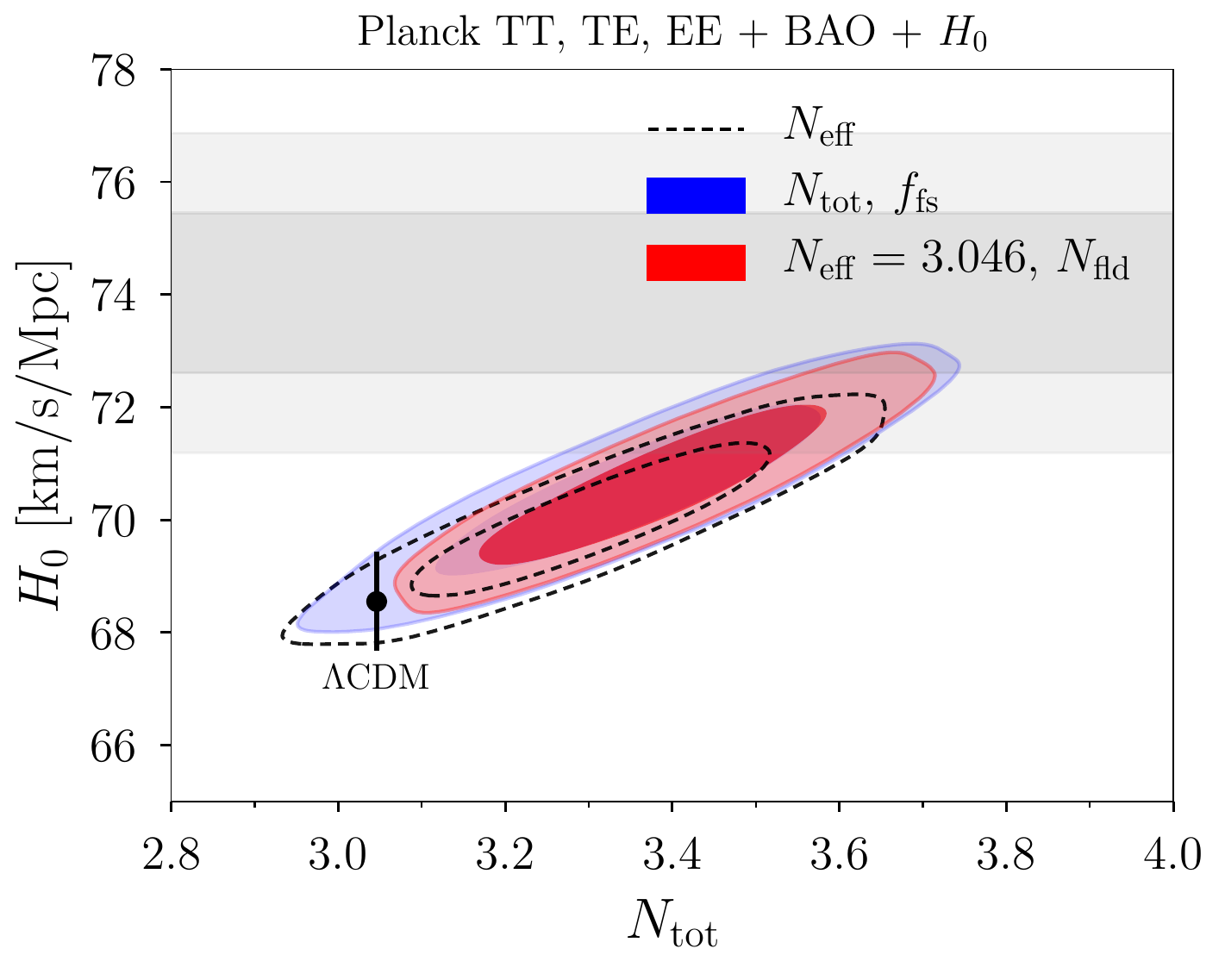}
    \caption{Marginalized posterior of the total radiation degrees of freedom $\Ntot$ and the present value of the Hubble rate, $H_0$, 
      in different models. In the left panel we combine Planck and BAO data, while in the right we add the distance ladder measurement of $H_0$~\cite{Riess:2019cxk}. 
		In both panels the gray horizontal band represents this measurement and the black dot is the $\Lambda$CDM value with $2\sigma$ uncertainties.
		The darker inner (lighter outer) regions correspond to 68\% (95\%) confidence regions.
		\label{fig:Ntot-H0}}
\end{figure}

\paragraph{Extra radiation and \texorpdfstring{$H_0$}{H0}}
An increase in the total amount of radiation, interacting or free-streaming, leads to an increase in $H_0$. This is easily understood from the fact that the extra radiation decreases the size of the sound horizon, as discussed in Sec.~\ref{sec:boltzmann}, which requires a decrease in the angular distance $D_A$ in order to keep the position of the peaks fixed. This decrease in $D_A$ is largely driven by an increase in $H_0$. The correlation between $\Ntot$ and $H_0$ can be readily seen in Fig.~\ref{fig:Ntot-H0} for all models where the radiation density is allowed to vary.

\begin{figure}
	\centering
	\includegraphics[width=0.5\textwidth]{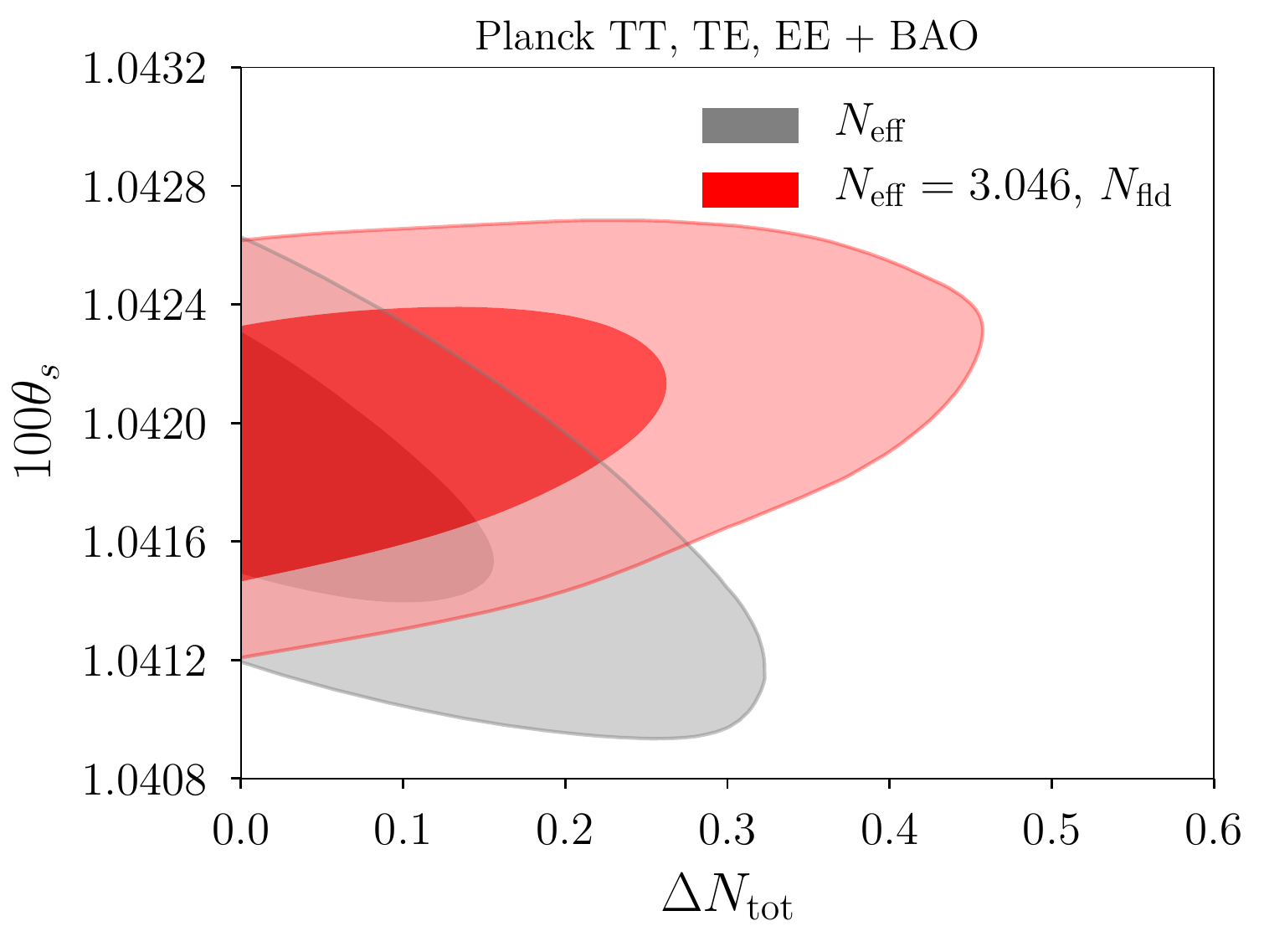}
    \caption{Marginalized posterior of the additional radiation degrees of freedom (either $\Delta\Ntot= \Neff - 3.046$, or $\Delta\Ntot = \Nfld$ with $\Neff = 3.046$) 
		and the angular scale of the sound horizon $\theta_s$ as derived from the Planck + BAO data sets. Free-streaming ($\Neff$) and non-free-streaming 
		($\Nfld$) radiation exhibit different correlation with $\theta_s$ due to their opposite effects on the phase shift of acoustic peaks in the CMB. 
		Supersonic propagation of free-streaming radiation perturbations shifts acoustic peaks to larger angular scales; in order to keep the physical peak locations 
		the same $\theta_s$ must be decreased. Conversely, non-free-streaming radiation reduces this phaseshift, requiring $\theta_s$ to increase to keep 
		peak location the same. The darker inner (lighter outer) regions correspond to 68\% (95\%) confidence regions.
	}
	\label{fig:Ntot-theta}
\end{figure}

\paragraph{Interacting versus free-streaming radiation}
The effect of the phase shift associated with free-streaming species can be directly seen in Fig.~\ref{fig:Ntot-theta}. From it one can easily see that in the $\Neff$ model an increase in the radiation density is anti-correlated with the angular size of the sound horizon. This is expected from the fact that the supersonic propagation of free-streaming radiation shifts the acoustic peaks of the CMB to larger scales. In order to maintain the location of the peak in the power spectrum fixed this requires a decrease in the physical size of the sound horizon angular scale $\theta_s = r_s/D_A$, as can be seen from Eq.~\ref{eq:phase-shift}. The effect of increasing the energy density in interacting radiation on the other hand decreases the phase shift associated with the neutrinos and thus requires an increase in the angular sound horizon scale.

\begin{figure}
	\centering
	\includegraphics[width=0.47\textwidth]{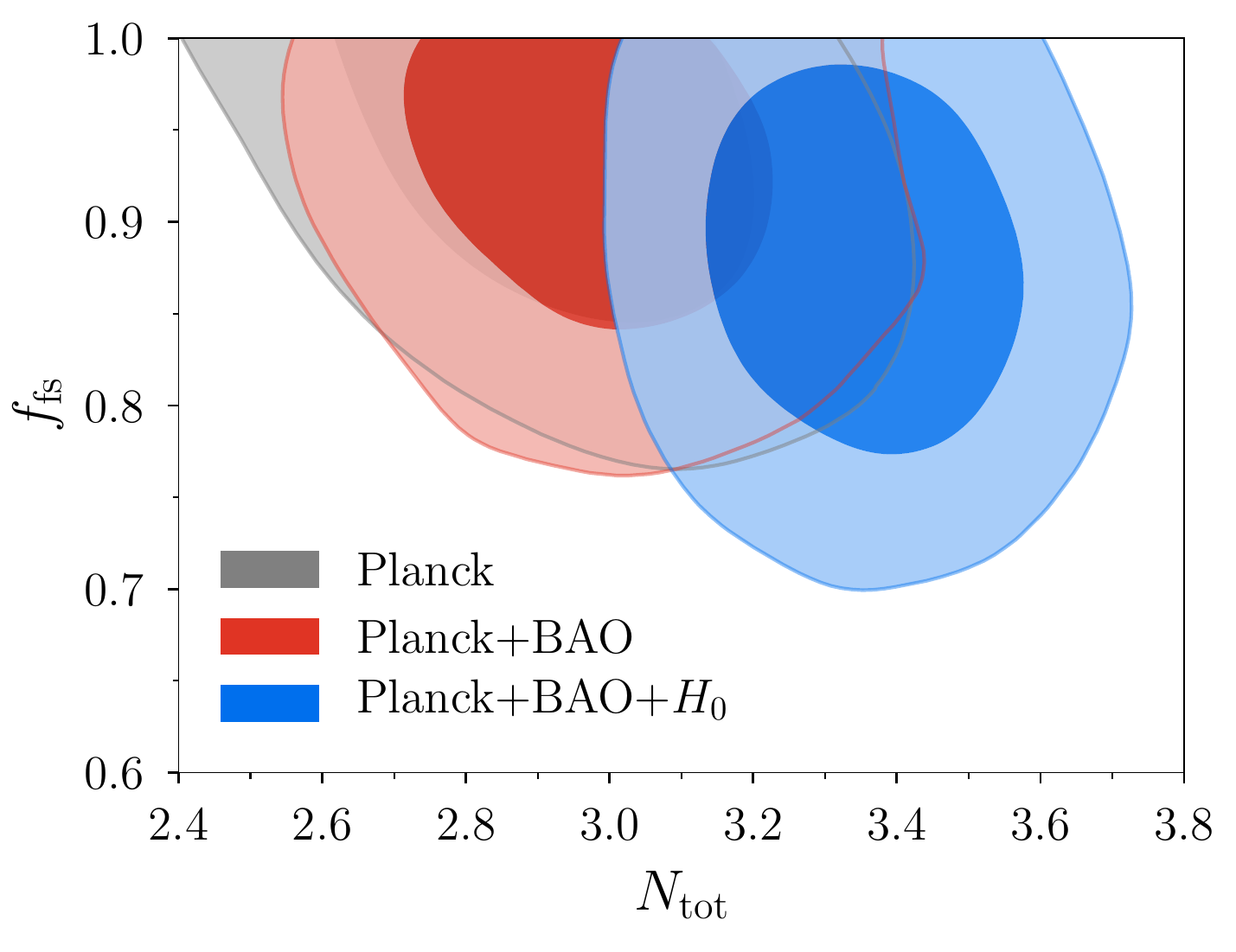}
	\includegraphics[width=0.47\textwidth]{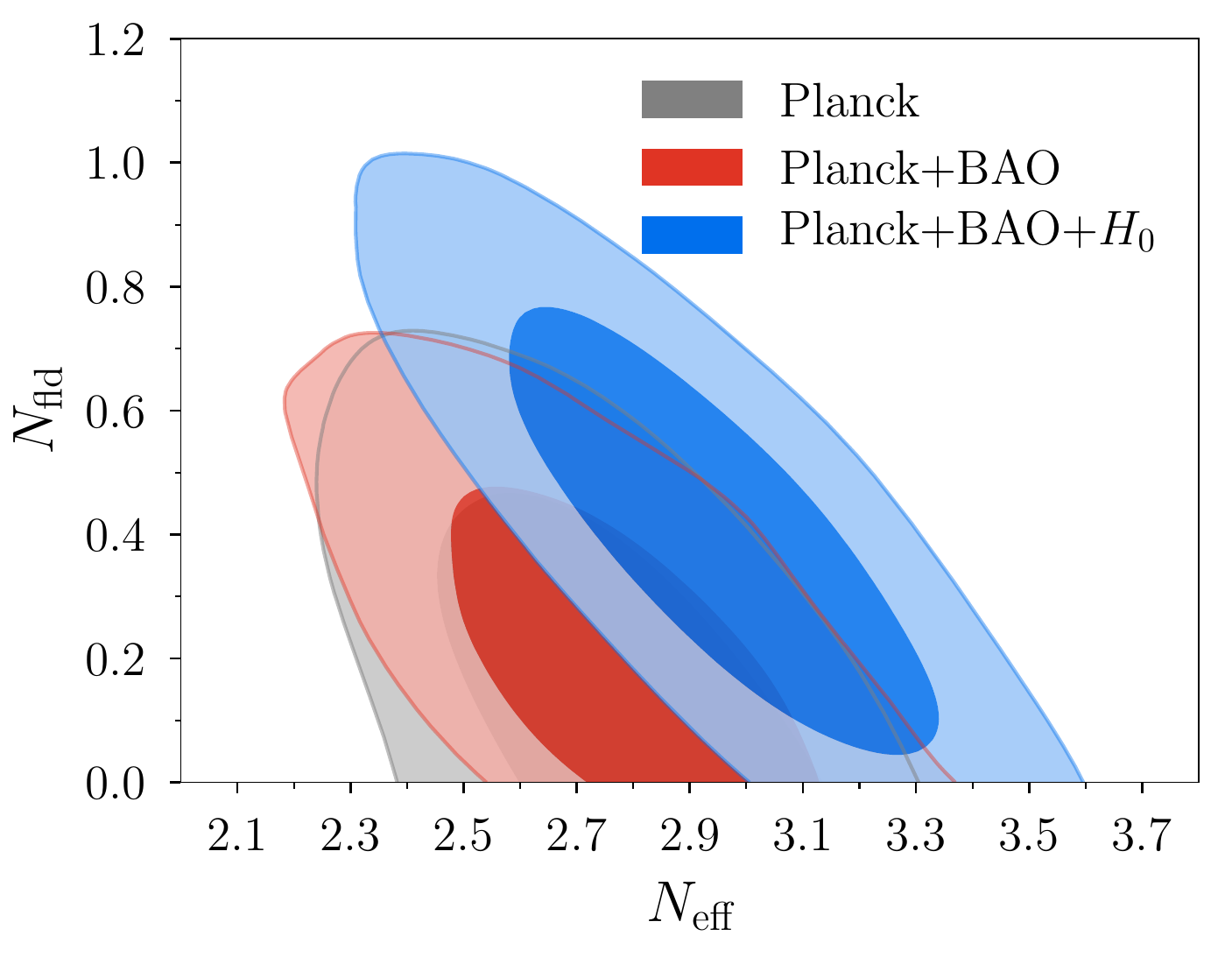}
    \caption{Marginalized posterior for the model with free-streaming and non-free-streaming species for different combinations of cosmological data. 
      In the left panel the posterior is shown as a function of the total radiation degrees of freedom $\Ntot$ and their free-streaming fraction $\ffs$, while 
      the right panel instead uses the number of free-streaming and non-free-streaming relativistic degrees of freedom, $\Neff$ and $\Nfld$, respectively.
    In all cases, the darker inner (lighter outer) regions correspond to 68\% (95\%) confidence regions.}
	\label{fig:ntot-fs}
\end{figure}

\begin{figure}
	\centering
%	\includegraphics[width=0.5\textwidth]{graphics/ffs_TTTEEE_lowE_lens_BAO}
%    \caption{Marginalized posteriors of the fraction of energy in free-streaming radiation, $\ffs$ in models with fixed and varying $\Ntot$ (solid black and dashed red lines, respectively) for the {\it Planck} TT, TE, EE + BAO data set. The dotted gray line denotes the 95.4\% CL lower limit on $\ffs$ in both models (which happen to coincide -- see Tabs.~\ref{tab:ntot_fixed_results} and~\ref{tab:ntot_planck_bao}).}
	\includegraphics[width=0.47\textwidth]{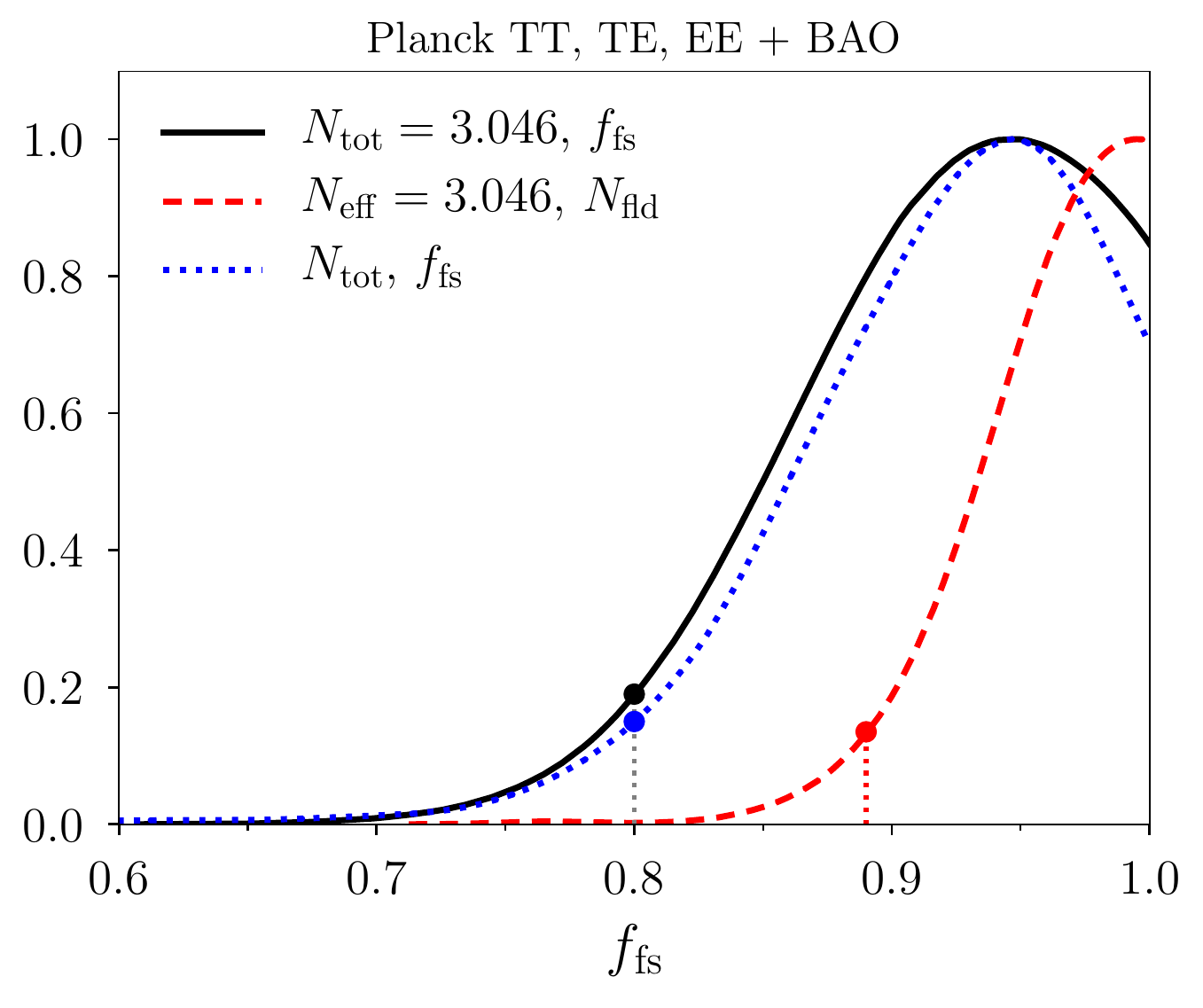}
	\includegraphics[width=0.47\textwidth]{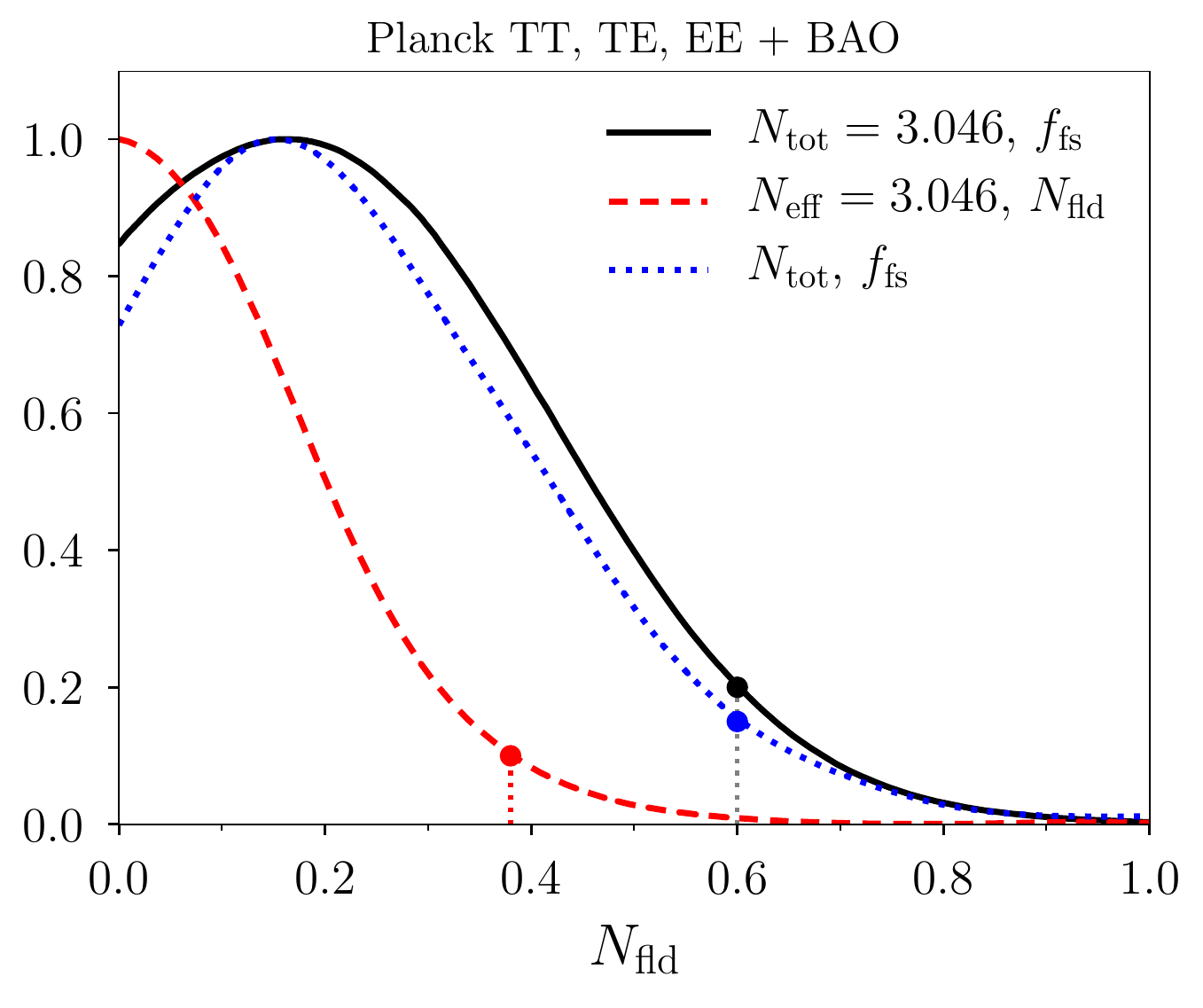}
    \caption{Marginalized posteriors of the fraction of energy in free-streaming radiation $\ffs$ (left panel) and the number 
      of non-free-streaming degrees of freedom $\Nfld$ (right panel) in models with fixed and varying $\Ntot$ (solid, and dashed or dotted lines, respectively) for the {\it Planck} TT, TE, EE + BAO data set. The solid dots denote the 95.4\% CL lower (upper) limit on $\ffs$ ($\Nfld$) in the left (right) panel. 
      These limits happen to coincide for the $(\Ntot=3.046,\ffs)$ and $(\Ntot, \ffs)$ models -- see Tabs.~\ref{tab:ntot_fixed_results} and~\ref{tab:ntot_planck_bao}.}
	\label{fig:ntot-fs-1d}
\end{figure}

\begin{figure}
  \centering
  \includegraphics[width=0.5\textwidth]{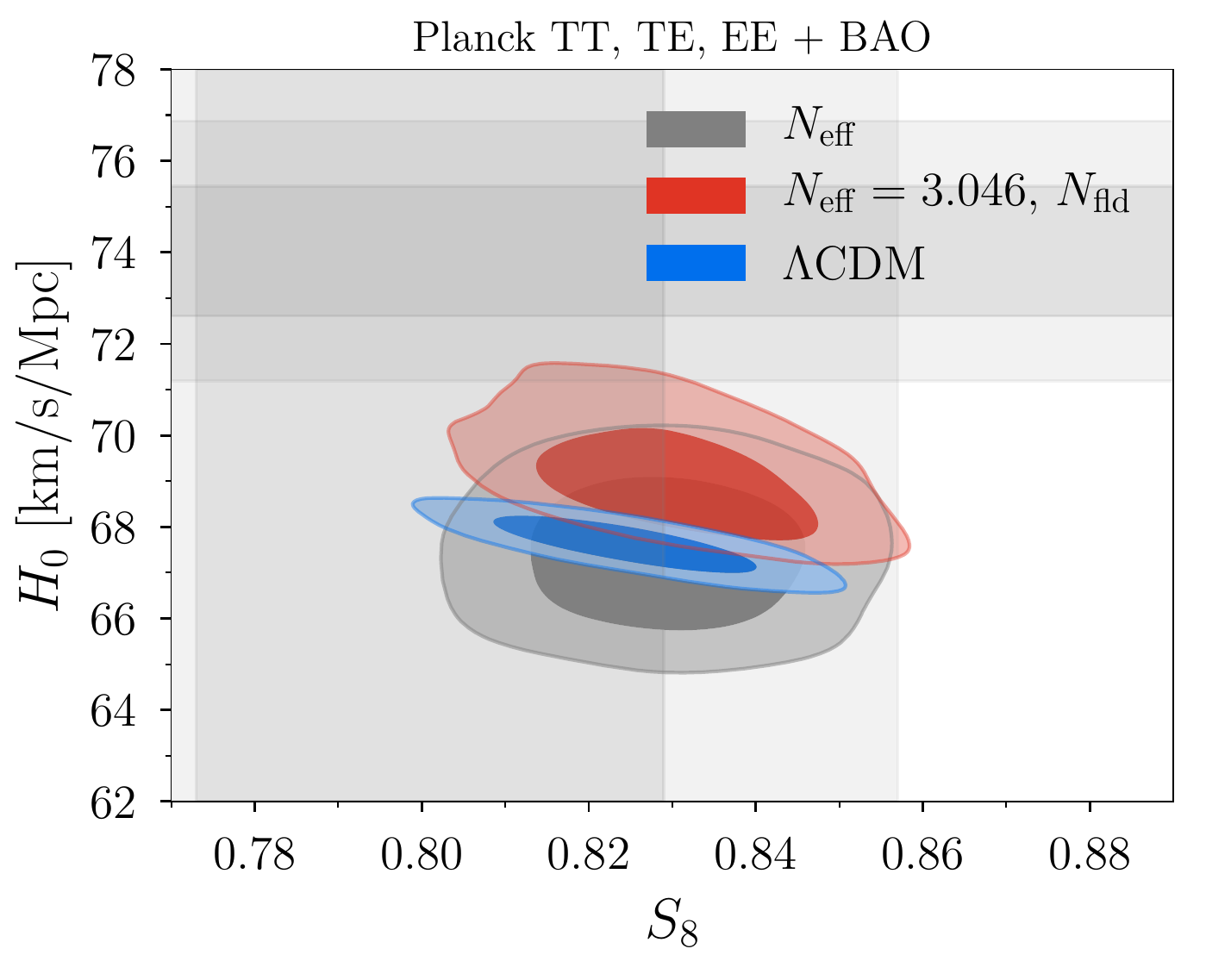}
  \caption{Marginalized posteriors of $H_0$ and $S_8= \sigma_8 \sqrt{\Omega_m/0.3}$ in different models (colored regions) for {\it Planck} TT, TE, EE + BAO data set and 
    the corresponding direct late-time measurements (gray bands) from Refs.~\cite{Riess:2019cxk} and~\cite{Abbott:2018wzc}. 
    The darker inner and lighter outer contours correspond to $68$\% and $95$\% confidence regions, 
respectively.
\label{fig:S8-H0}}
\end{figure}

\paragraph{Implications for the Hubble and \texorpdfstring{$\sigma_8$}{sigma8} Tensions}
The fraction of interacting radiation allowed by the data is not significantly
increased by letting $\Ntot$ vary as seen in Figs.~\ref{fig:ntot-fs}
and~\ref{fig:ntot-fs-1d}. The only exception is when the likelihood for the
local $H_0$ measurements is included in the scan, which brings the allowed
range of $\Ntot$ to slightly larger values and leads to a mild preference for a
non-zero fraction of interacting radiation. This can be partially explained by
recalling that the increase in $\Ntot$ required to accommodate a larger $H_0$
leads to an increase in the ratio between the photon diffusion length and the
sound horizon scale. In addition, the presence of free-streaming species
decreases the gravitational potential at horizon crossing, and therefore leads
to a smaller radiation driving effect for modes that enter the horizon before
matter radiation equality. Both of these effects decrease the temperature power
spectrum at larger $\ell$. Allowing part of the radiation to be interacting
increases the size of the gravitational potential at horizon crossing (see
Eq.~\ref{eq:psi_ic}) in comparison to models with only $\Neff$, and therefore
leads to less suppression of power at high $\ell$. This effect enables 
the data to tolerate slightly larger values of $\Ntot$ as shown in Fig.~\ref{fig:ntot-fs}. 
This, in turn, leads to a larger overlap of the Planck+BAO $H_0$ posterior with the local 
measurement compared to free-streaming radiation, which is illustrated in Fig.~\ref{fig:Ntot-H0}.
The same effect also explains why
models with interacting radiation lead to smaller values for $n_s$ compared to
free-streaming only models as can be seen in Tab.~\ref{tab:ntot_planck_bao_H0}.
The smaller value for $n_s$ leads to a smaller $\sigma_8$ for models with
interacting radiation in comparison to models with only free-streaming species.
However, as shown in Fig.~\ref{fig:S8-H0}, when looking at $S_8 = \sigma_8\sqrt{\Omega_m/0.3}$ (a quantity more directly related to measurements), both
models are within the $2\sigma$ allowed region of low-$z$ measurement of Ref.~\cite{Abbott:2018wzc}, $S_8 = 0.801^{+0.028}_{-0.026}$.

In Table~\ref{tab:delta_chi2_H0} we break down the $\chi^2$ for the best fit points to the {\it Planck} TT, TE, EE + BAO + $H_0$ likelihood for each model by the different likelihoods used. We find that for all of the models the improvement in $\chi^2$ is driven almost exclusively by the improved fit to the $H_0$ likelihood. In particular all models provide a worse fit to the CMB power spectrum, except at low-$\ell$ ($\ell \leq 29$), where they improve the temperature power spectrum and for the models with $\Nfld$ there is also improvement for low-$\ell$ polarization. 

\begin{table}
\centering
  \begin{tabular}{|l|c|c|c|c|}
\hline
Data Set &  $\Neff$  &  $\Neff = 3.046,\; \Nfld$  &  $\Ntot,\;\ffs$ \\
\hline
TTTEEE  &    $+2.68$  &  $+6.24$  &  $+6.24$\\
low-$\ell$ TT  &  $-0.63$  &  $-0.56$  &  $-0.56$\\
low-$\ell$ EE  &  $+0.09$  &  $-1.06$  &  $-0.29$\\
lensing  &  $+0.17$  &  $+0.8$  &  $+0.39$\\
BAO     &  $+0.39$  &  $+0.73$  &  $+1.04$\\
$H_0$   &  $-4.99$  &  $-9.93$  &  $-10.81$\\
\hline
total   &  $-2.3$   &  $-3.81$  &  $-4.02$\\
\hline
  \end{tabular}
  \caption{Best fit $\Delta \chi^2 = \chi^2- \chi^2_{\Lambda\mathrm{CDM}}$ for models with extra radiation species. 
 In each case the best fit points correspond to the full Planck + BAO + $H_0$ data set combination of likelihoods.
 Low-$\ell$ refers to the $\ell \leq 29$ Planck likelihoods. 
 \label{tab:delta_chi2_H0}
}
\end{table}

\begin{figure}
	\centering
	\includegraphics[width=0.5\textwidth]{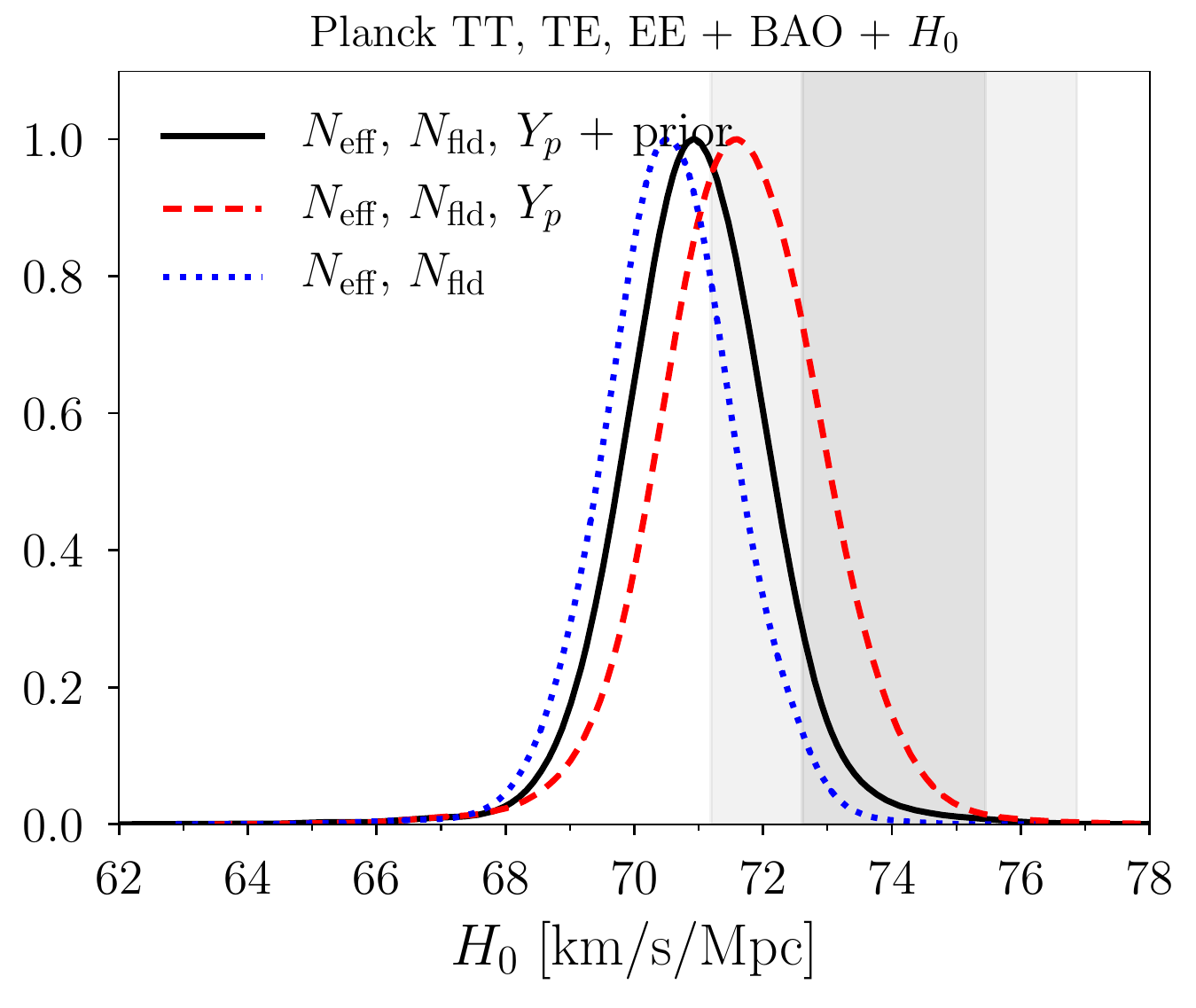}
    \caption{Marginalized posteriors of $H_0$ for the $(\Neff, \, \Nfld)$ model with $Y_p$ fixed by BBN (blue dotted), or 
      allowed to vary disregarding direct measurements (red dashed) and including direct measurements (black solid). 
      The gray band shows the local $H_0$ measurement~\cite{Riess:2019cxk}, with darker inner and lighter outer contours corresponding to $1\sigma$ and $2\sigma$ confidence regions, 
respectively.
    \label{fig:yp1d}}
\end{figure}

\begin{figure}
	\centering
	\includegraphics[width=0.5\textwidth]{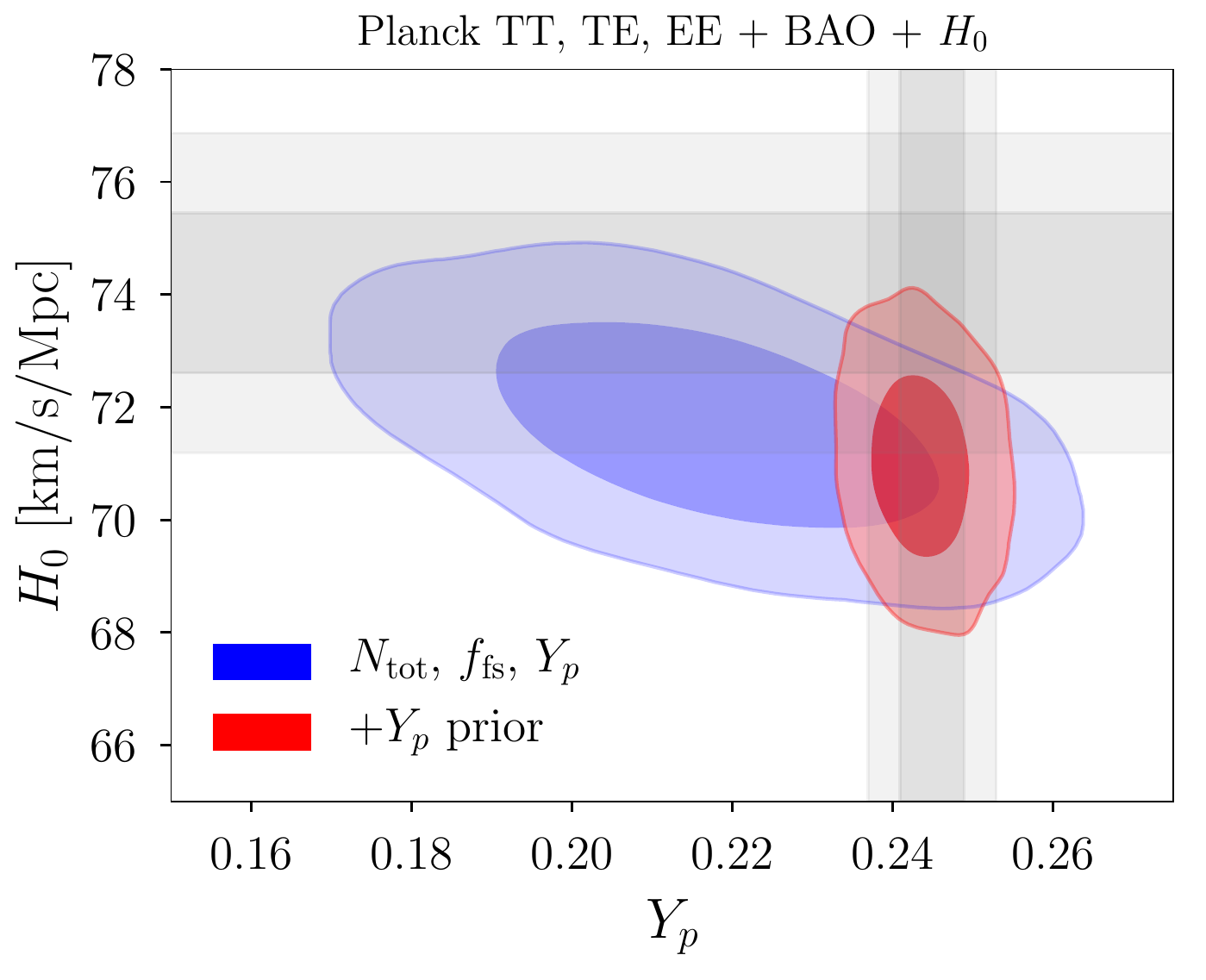}
	\caption{Marginalized posteriors of $H_0$ and $Y_p$ for the $(\Neff, \, \Nfld)$ model with $Y_p$ allowed to vary disregarding direct measurements (blue) and including direct measurements as a prior (red). 
      The gray bands show the direct late-time measurements of $H_0$~\cite{Riess:2019cxk} (horizontal band) and $Y_p$~\cite{Aver:2015iza} (vertical band), with darker 
      inner and lighter outer contours corresponding to $1\sigma$ and $2\sigma$ confidence regions, 
respectively.
	\label{fig:yp2d}}
\end{figure}

\paragraph{Variation of primordial helium fraction \texorpdfstring{$Y_p$}{Yp}}
It is well known that there is a partial degeneracy between $\Ntot$ and $Y_p$ due to their effects on the diffusion damping scale (see the discussion around Eq.~\ref{eq:damping_scale} and, e.g., Ref.~\cite{Hou:2011ec} for a clear explanation). 
In order to see the maximum effect of $Y_p$ in reducing the Hubble tension we considered models with freely varying $\Ntot$ and $Y_p$ (i.e., $Y_p$ is \emph{not} calculated from BBN). The results for those runs are shown in Table~\ref{tab:Yp_vary_results}. One sees that the extra freedom to adjust $Y_p$ significantly increases the allowed range for $\Ntot$ and consequently also for $H_0$. These results are relevant to cosmological scenarios where $Y_p$ can vary independently of $\Ntot$, 
which can occur in some of the models described in Sec.~\ref{sec:models}. However, aside from the BBN consistency requirement, there are also direct measurements of the primordial ${}^4$He abundance which place constraints on $Y_p$ that only depend on late Universe considerations. We have explored the effects of including such measurements in cosmological fits as an additional Gaussian prior with $Y_p = 0.2449\pm 0.0040$~\cite{Aver:2015iza}. We compare the results for runs including a $Y_p$ prior to those without in Figs.~\ref{fig:yp1d} and~\ref{fig:yp2d}. One can directly see from the figures that the inclusion of the local measurements as a prior leads to very similar results to using BBN to determine $Y_p$, and therefore does not significantly improve the fit to local $H_0$ measurements.

\begin{table}
	\centering
	\begin{tabular}{|l|c|c|c|} 
      \hline
      & TT, TE, EE  &  TT, TE, EE + BAO & TT, TE, EE + BAO + $H_0$  \\
\hline
$100\theta_s$  &  $1.0428^{+0.0011}_{-0.001}$  &  $1.0428^{+0.00093}_{-0.00102}$  &  $1.0417^{+0.00085}_{-0.00106}$\\
$100\Omega_b h^2$  &  $2.24^{+0.023}_{-0.026}$  &  $2.241^{+0.019}_{-0.021}$      &  $2.26^{+0.019}_{-0.019}$\\
$\Omega_c h^2$    &  $0.1201^{+0.0049}_{-0.0062}$  &  $0.1198^{+0.0043}_{-0.0052}$  &  $0.1295^{+0.0044}_{-0.0046}$\\
$\ln 10^{10}A_s$  &  $3.026^{+0.019}_{-0.02}$  &  $3.027^{+0.019}_{-0.017}$       &  $3.033^{+0.02}_{-0.019}$\\
$n_s$           &  $0.9565^{+0.0089}_{-0.0086}$  &  $0.9572^{+0.0084}_{-0.0075}$  &  $0.9613^{+0.0091}_{-0.0079}$\\
$\tau$         &  $0.0547^{+0.0074}_{-0.0081}$  &  $0.0551^{+0.0067}_{-0.0074}$   &  $0.0551^{+0.0072}_{-0.0079}$\\
$\Ntot$         &  $3.05^{+0.33}_{-0.41}$      &  $3.04^{+0.26}_{-0.33}$          &  $3.66^{+0.26}_{-0.26}$\\
%$\ffs$          &  $0.905^{+0.095}_{-0.025}$   &  $0.908^{+0.092}_{-0.025}$       &  $0.867^{+0.078}_{-0.059}$\\
$\ffs$          &  $>0.78$   &  $>0.79$       &  $0.867^{+0.078}_{-0.059}$\\
$Y_p$              &  $0.241^{+0.022}_{-0.019}$  &  $0.241^{+0.02}_{-0.018}$      &  $0.217^{+0.019}_{-0.019}$\\
\hline
$H_0$ [km/s/Mpc]              &  $68.2^{+2.0}_{-2.7}$    &  $68.2^{+1.5}_{-1.8}$            &  $71.6^{+1.3}_{-1.2}$\\
$r_s^\mathrm{drag}$ [Mpc]         &  $147.1^{+3.6}_{-3.2}$  &  $147.2^{+2.8}_{-2.7}$   &  $141.7^{+2.2}_{-2.2}$\\
$\sigma_8$          &  $0.815^{+0.011}_{-0.012}$  &  $0.815^{+0.011}_{-0.011}$    &  $0.829^{+0.011}_{-0.01}$\\
\hline
$\chi^2_{\mathrm{tot}}$ & 2774.55  &  2781.46  &  2788.85 \\
\hline
	\end{tabular} 
    \caption{Results for a model with $\Ntot$, $\ffs$ and varying $Y_p$ (disregarding its direct measurement). 
      We present the marginalized mean $\pm 1 \, \sigma$ error for all the parameters, with the exception of $\ffs$ for which we show the $95\% $ lower limit.
	\label{tab:Yp_vary_results}}
\end{table}

\subsection{Particle Physics Interpretation}
\begin{figure}
	\centering
	\includegraphics[width=0.47\textwidth]{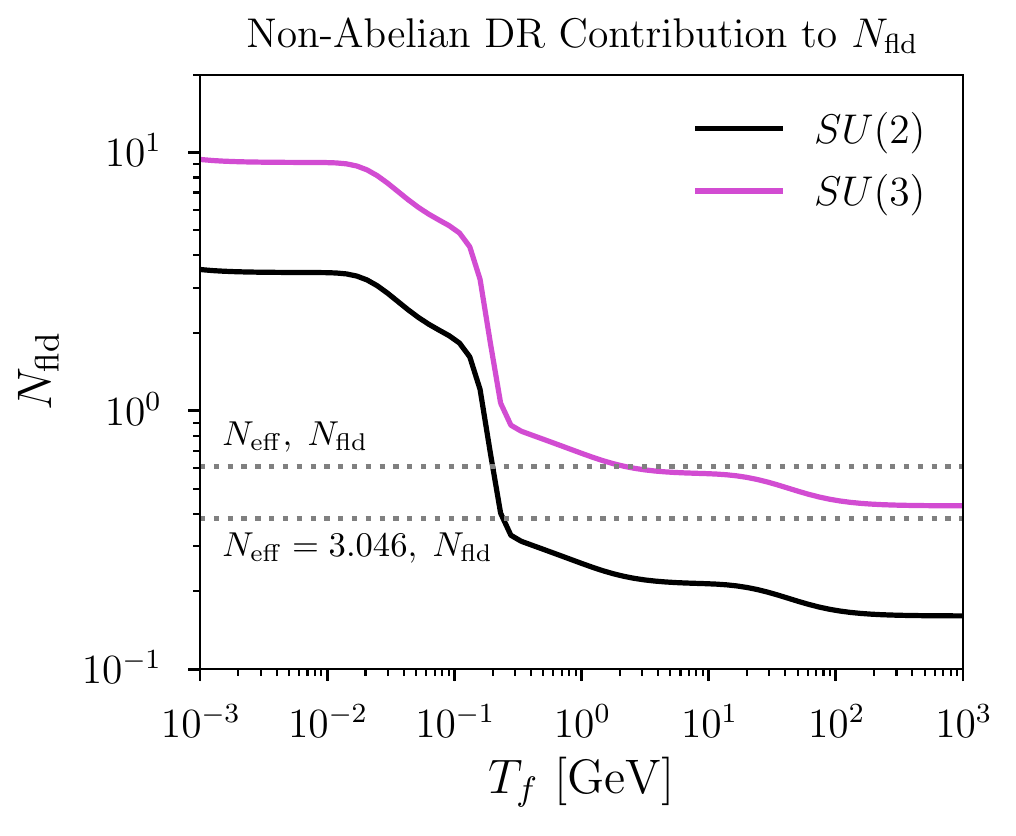}
	\includegraphics[width=0.47\textwidth]{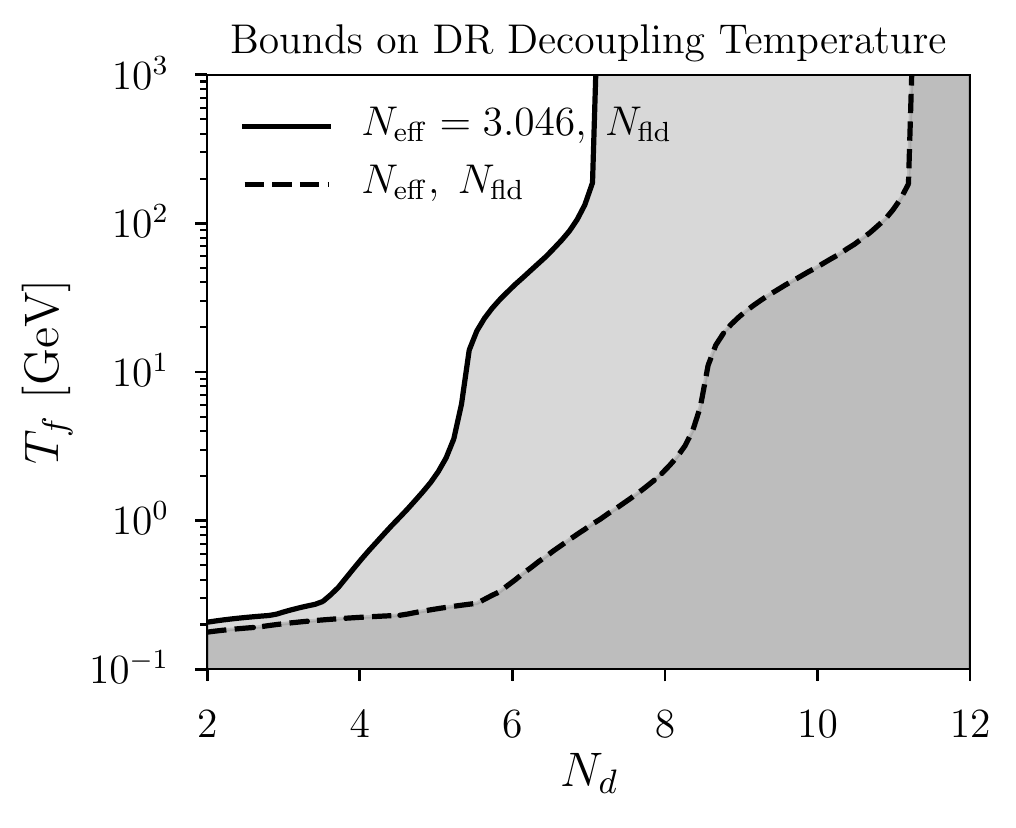}
    \caption{Constraints on non-Abelian dark radiation that decoupled from the SM at temperature $T_f$. In the left 
    panel we show the predictions for $\Nfld$ as a function of $T_f$ for DR from an $SU(2)$ ($SU(3)$) dark sector as 
    the lower (upper) solid line. Upper bounds on $\Nfld$ from $(\Neff,\;\Nfld)$ and $(\Neff = 3.046,\Nfld)$ scans using Planck+BAO 
    data are shown as dotted gray lines. In the right panel we translate these constraints into bounds on 
    $T_f$ as a function of the number of dark gluons $N_d$ (e.g., $N_d = 3$ in $SU(2)$). Gray shaded region is excluded at 95\% C.L.
    \label{fig:na_dr_constraints}}
\end{figure}

In this section we interpret our results in terms of the illustrative models described in Sec.~\ref{sec:models}.
\paragraph{Non-Abelian Dark Radiation}
First, we translate the constraints on $\Nfld$ into an upper bound on the number of dark gluons $N_d$ and the ratio of DR to SM temperature 
using Eq.~\ref{eq:nfld_from_dr}
\beq
N_d \left(\frac{T_d}{T_\gamma}\right)^4 \leq 
\begin{cases}
  0.087 & 95\%\; \mathrm{C.L.} \;(\Neff=3.046,\;\Nfld)\\
  0.14 & 95\%\; \mathrm{C.L.}\; (\Neff,\;\Nfld),
\end{cases}
\label{eq:na_dr_constraint_generic}
\eeq
where we used the Planck+BAO results from Tab.~\ref{tab:ntot_planck_bao} and Fig.~\ref{fig:ntot-fs-1d}.
We see that allowing both $\Neff$ and $\Nfld$ to vary leads to somewhat weaker constraints. Note that this 
result does not assume that the DR and SM sectors were ever in thermal equilibrium.
However, if there exists a  coupling between the non-Abelian dark sector, it must necessarily be through a 
higher dimensional operator. In this case it is natural for the SM and DS to be in equilibrium at early times 
but decoupled at late times, leading to a prediction for the temperature ratio $T_d/T_\gamma$ given 
in Eq.~\ref{eq:decoupled_ds_temp}. This prediction only depends on the SM-DR decoupling temperature $T_f$, and the 
assumption that there was no non-SM entropy injection between $T_f$ and today. In Fig.~\ref{fig:na_dr_constraints} 
we show the constraints implied by Eq.~\ref{eq:na_dr_constraint_generic} and the early-time equilibrium hypothesis. 
In the left panel the prediction of Eq.~\ref{eq:nfld_from_dr} for two gauge groups is shown as a function of $T_f$ 
illustrating the constraining power of cosmological data. The strongest cosmological constraints, for example, 
rule out DR from an $SU(3)$ sector that was in equilibrium with the SM at \emph{any} temperature. 
In the right panel of Fig.~\ref{fig:na_dr_constraints} we show the lower bound on $T_f$ as a function of the 
number of dark gluons $N_d$. The strongest Planck and BAO constraints therefore exclude the possibility of a non-free-streaming dark sector with $N_d > 7$ 
being in thermal equilibrium with the SM at any temperature (again, assuming no non-standard entropy injections after $T_f$).

\paragraph{Late Equilibration of a Dark Sector}
Our results can be used to constrain several limits of the late equilibration scenario described in Sec.~\ref{sec:late_equilibration}.
The most straightforward application is the case when a neutrino-coupled mediator $\phi$
equilibrates with one or more neutrinos well before modes relevant for the CMB enter the horizon; this occurs if 
\beq
\sqrt{\lambda_\nu} m_\phi \gtrsim 6\times 10^{-6}\;\eV,
\eeq
where we used Eq.~\ref{eq:ds_nfs_temp}, taking $T^{\mathrm{nfs}}_\phi \gtrsim 20 (4/11)^{1/3}\;\eV$ (roughly corresponding to the 
time when modes with $k \sim 0.2/\mathrm{Mpc}$ enter the horizon). 
The equilibrium must persist until well after matter-radiation equality, requiring $m_\phi \ll \eV$.\footnote{If $m_\phi \gg \eV$, the 
mediator goes non-relativistic and heats the neutrinos relative to photons, increasing $\Neff$ without changing neutrino free-streaming 
during the CMB epoch. Bounds on $\Delta\Neff$ can be applied in this regime. If the non-relativistic transition occurs during the CMB era, 
a different analysis is required since the equation of state of the $\phi-\nu$ fluid and phase-space distributions have non-trivial evolution.} 
These conditions ensure that $\Ntot \approx 3$
and that the $\phi-\nu$ bath behaves as a relativistic fluid throughout the CMB era. This scenario then directly maps onto 
the $(\Ntot=3.046,\ffs)$ scan, with the naive expectation that at least one neutrino species is not free-streaming.
In Fig.~\ref{fig:ntot-fixed-ffs-1d} and Tab.~\ref{tab:ntot_fixed_results} we found that the free-streaming 
fraction has to be larger than 0.8 at 95\% C.L. for the CMB+BAO dataset, robustly excluding this part of the late equilibration parameter space.

\begin{figure}
	\centering
	\includegraphics[width=0.5\textwidth]{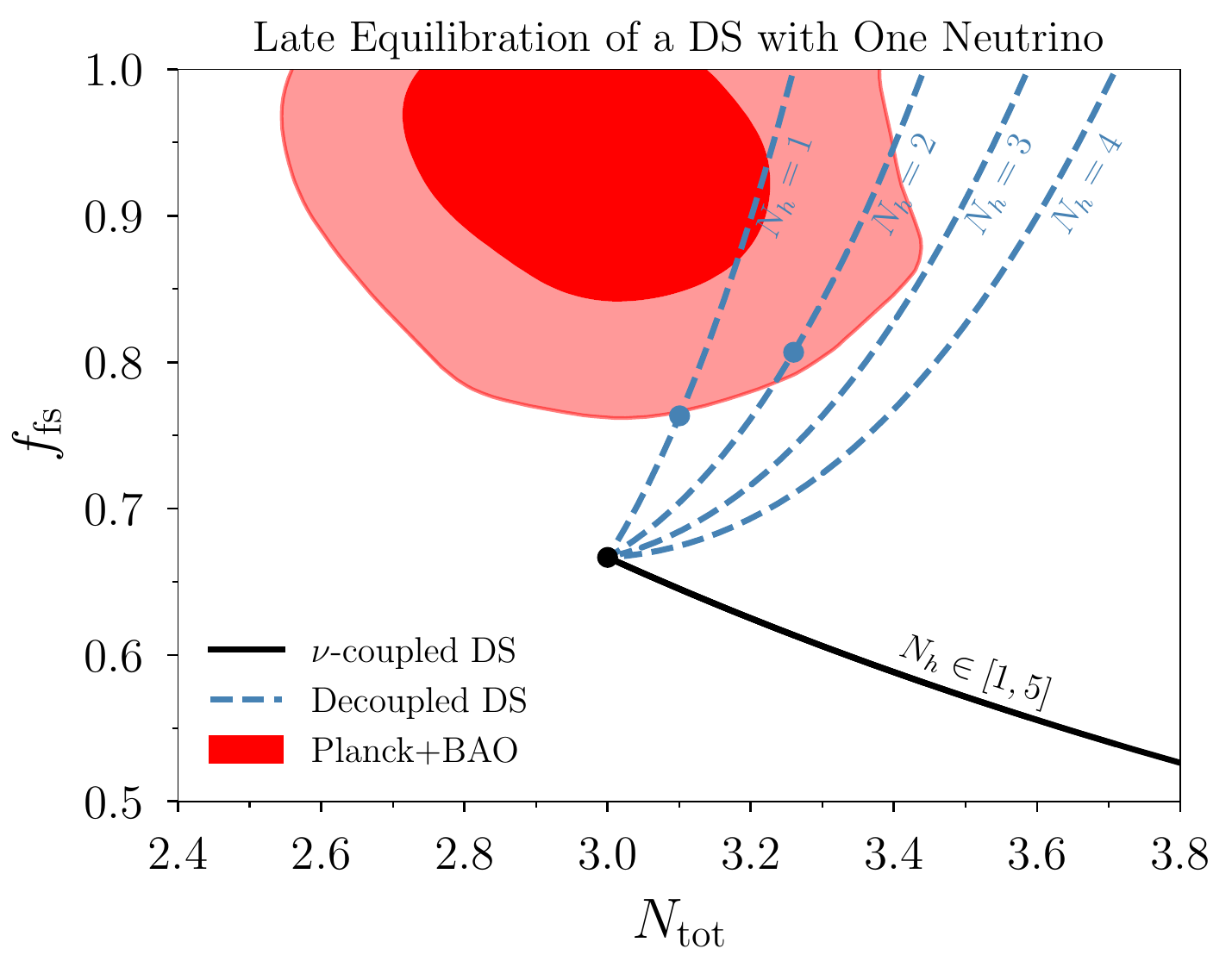}
    \caption{Predictions of models with late equilibration of a dark sector (DS) that couples to a single neutrino mass eigenstate. The black line shows the 
      case where the DS remains coupled to that neutrino throughout the CMB epoch, so that the only free-streaming species are the 
      remaining two neutrinos. This contour is obtained by varying the total number of DS states $N^\prime$ and the number of ``heavy'' 
      states $N_h$ that become non-relativistic after equilibration but before the CMB epoch. Larger $N_h$ results in a higher density 
      of non-free-streaming states, thereby lowering $\ffs$. The dashed blue lines show the predictions of an alternative scenario where after equilibration, 
      $N_h$ DS states become non-relativistic, including the mediator responsible for $\nu$-DS interaction, thereby decoupling the two sectors 
      once again. This heats both the DS states and the single neutrino eigenstate; now, however, all neutrinos contribute to the free-streaming fraction, including 
      the one that used to interact with the DS. The dots on the $N_h = 1$ and $2$ lines correspond to $N^\prime = 3$, with larger $N^\prime$ falling below the dots.
      The Planck+BAO posterior distribution (see Fig.~\ref{fig:ntot-fs}) is shown in red with 
      the darker inner (lighter outer) regions corresponding to 68\% (95\%) confidence regions.
    \label{fig:late_eq_constraints}}
\end{figure}

The second possibility highlighted in Sec.~\ref{sec:late_equilibration} was that $\phi$ is a portal to a larger dark sector (with a total number $N^\prime$ 
degrees of freedom), so late equilibration also populates these additional states. If some number $N_h$ of these particles are ``heavy'', i.e. 
they become non-relativistic before the CMB era, they will heat the remaining interacting particles compared to the SM, 
which increases $\Ntot$. There are two interesting limits of this scenario. First, we assume that the remaining DS states stay in 
equilibrium with neutrinos throughout the CMB era. Then, the only contributions to the free-streaming fraction $\ffs$ come from 
those neutrino mass eigenstates that do not interact with the DS:
\beq
\ffs \approx \frac{3 - N_\nu}{\Ntot},
\label{eq:nu_coupled_ds}
\eeq
where $N_\nu$ is the number of interacting neutrinos and $\Ntot$ is given by Eq.~\ref{eq:ntot_from_late_equilibration}.
This scenario is generalization of the one described in the previous paragraph, which allows for $\Ntot > 3$. The 
prediction this model is shown in Fig.~\ref{fig:late_eq_constraints} for $N_\nu = 1$, varying $N^\prime$ and different choices of $N_h$. 
Note that since the numerator of Eq.~\ref{eq:nu_coupled_ds} only depends on $N_\nu$, all choices of $N_h$ collapse onto the same contour.
It is also clear from this equation that we expect $\ffs < 2/3$ for this scenario, which is robustly excluded by cosmological data.

The second interesting limit corresponds to the case where the $N_h$ heavy states include the mediator $\phi$. 
When these particles become non-relativistic, they heat both the interacting neutrino and the light DS states; however, 
after this point the neutrinos and the DS are decoupled. In this case, all neutrino eigenstates contribute to $\ffs$, but 
$N_\nu$ of them have a distinct temperature; we find that 
\beq
\ffs \approx \frac{1}{\Ntot}\left[3 - N_\nu + N_\nu \left(\frac{N_\nu}{N_\nu + N^\prime}\right)\left(\frac{N_\nu+N^\prime}{N_\nu + N^\prime - N_h}\right)^{4/3}\right],
\eeq
where $\Ntot$ is given in Eq.~\ref{eq:ntot_from_late_equilibration}. The predictions for 
this scenario are shown by blue dashed lines in Fig.~\ref{fig:late_eq_constraints} for $N_\nu = 1$. In contrast to the previous case, 
the temperature of the now-decoupled $N_\nu$ eigenstates depends on $N_h$, which ensures that different choices of this quantity 
give different predictions in the $\Ntot-\ffs$ plane. In all cases, however, for $N^\prime\gg N_\nu,\;N_h$ the contribution of 
the $N_\nu$ neutrinos to $\ffs$ is diluted as the energy is shared amongst many DS states, and $\ffs$ approaches $2/3$ (i.e. only 
the non-interacting $3-N_\nu$ neutrinos contribute to the free-streaming fraction as before). In the other limit $N^\prime \sim N_h$ 
$\ffs$ approaches $1$ - the only remaining light relics are neutrinos with a slightly higher 
total energy density, resulting in $\Ntot > 3$. We see in Fig.~\ref{fig:late_eq_constraints} that cosmological data 
is able to distinguish between these different scenarios and place severe constraints on models with $N^\prime > 3$ or $N_h > 2$. 
We emphasize that these constraints apply as long as the equilibration and non-relativistic transitions occur 
outside of the CMB era. If this is not the case, then the various fluids have time-dependent equations of state and 
free-streaming behavior which is not captured by the simple cosmological model described in Sec.~\ref{sec:boltzmann}.

\section{Conclusion}
\label{sec:conclusion}

In this paper we have studied the status of cosmological models with extra radiation after Planck's final data release, allowing part of that radiation to be non-free-streaming. We also considered combinations of Planck's results with BAO and direct measurements of $H_0$.

When restricting the total radiation to be equal to that predicted by $\Lambda$CDM, $\Ntot = 3.046$, we find that Planck data alone requires the fraction of free-streaming radiation to be larger than $80 \%$. This shows that none of the neutrino flavors can have non-negligible interactions during the CMB epoch.

If the total amount of radiation is allowed to vary, we find that models which have an interacting component allow for slightly larger values of $\Ntot$, and therefore larger values for $H_0$. This reduces the tension between local measurements of $H_0$ and the CMB. Both the model with only extra interacting components and the model in which both free-streaming and non-free-streaming components are allowed to vary improve the total $\chi^2$ compared to the $\Neff$ case. However, if one takes into consideration the number of extra degrees of freedom in each model, $(\Neff=3.046,\Nfld)$ is the one most favored by data.

Interacting radiation models also lead to slightly lower values for $n_s$ when compared to free-streaming radiation models with the same total amount of radiation. This leads to smaller values for $\sigma_8$ which results in less tension between solutions to the $H_0$ tension and matter power spectrum measurements.

We also showed that the cosmological data is able to distinguish and constrain interesting particle physics models of interacting dark 
radiation, including non-Abelian dark sectors and late equilibration of neutrino-coupled particles.
In these models we focused on the parameter space where self-interactions within the fluid component are always in equilibrium. 
In many interesting scenarios these reactions fall out of or enter equilibrium -- see, e.g., Refs.~\cite{Forastieri:2019cuf,Kreisch:2019yzn}. 
If this occurs during the CMB era, it can leave a characteristic imprint on the power spectrum, possibly 
allowing further improvements to the fit of all cosmological data. These decoupling (or re-coupling) effects have been studied within the context of specific models relating to neutrino self-interactions. It would be interesting to investigate the impact of the final Planck data release on these and more general frameworks such as that of Ref.~\cite{Choi:2018gho}.

\section*{Acknowledgments}
We thank Francis-Yan Cyr-Racine, Deanna Hooper, Vivian Poulin, David E. Kaplan, Jose Bernal, Graeme Addison and Marilena LoVerde for useful conversations. 
We thank Yuhsin Tsai for useful discussions and comments on the draft.
This manuscript has been authored by Fermi Research Alliance, LLC under 
Contract No. DE-AC02-07CH11359 with the U.S. Department of Energy, Office of Science, Office of High Energy Physics. The research of GMT was supported in part by the NSF under Grant No. PHY-1914480, PHY-1914731 and by the Maryland Center for Fundamental Physics (MCFP).
We thank the Galileo Galilei Institute for Theoretical Physics for the hospitality and the INFN for partial support during the completion of this work. GMT acknowledges the KITP where part of this work was completed, supported in part by the National Science Foundation under Grant No. NSF PHY-1748958. We used \texttt{GetDist}~\cite{Lewis:2019xzd} to generate the figures in this work.

\appendix

\section{{\it Planck} TT results}
\label{sec:tt_only_results}

The constraints using only the {\it Planck} TT likelihood for the models with varying $\Ntot$ are given in Tab.~\ref{tab:ntot_planck_TT_only}.

\begin{table}
	\centering
	\begin{tabular}{|l|c|c|c|c|}
      \hline
 &  $\Lambda$CDM  &  $\Neff$  &  $\Neff=3.046$, $\Nfld$  &  $\Ntot$, $\ffs$ \\
\hline
$100\theta_s$  &  $1.0418^{+0.00044}_{-0.00043}$  &  $1.042^{+0.00068}_{-0.00068}$  &  $1.042^{+0.00045}_{-0.00045}$  &  $1.0429^{+0.00085}_{-0.00118}$\\
$100\Omega_b h^2$  &  $2.214^{+0.022}_{-0.021}$   &  $2.209^{+0.031}_{-0.032}$      &  $2.232^{+0.024}_{-0.028}$      &  $2.219^{+0.033}_{-0.031}$\\
$\Omega_c h^2$    &  $0.12^{+0.0016}_{-0.0016}$   &  $0.1188^{+0.0038}_{-0.0043}$   &  $0.1223^{+0.0019}_{-0.0028}$   &  $0.1196^{+0.0039}_{-0.0039}$\\
$\ln 10^{10}A_s$  &  $3.039^{+0.014}_{-0.015}$    &  $3.035^{+0.022}_{-0.02}$       &  $3.038^{+0.015}_{-0.015}$      &  $3.019^{+0.027}_{-0.022}$\\
$n_s$           &  $0.9636^{+0.0049}_{-0.005}$    &  $0.96^{+0.012}_{-0.013}$       &  $0.9659^{+0.0054}_{-0.0049}$   &  $0.953^{+0.014}_{-0.014}$\\
$\tau$         &  $0.0522^{+0.0077}_{-0.0076}$    &  $0.0518^{+0.0083}_{-0.008}$    &  $0.0532^{+0.0081}_{-0.0084}$   &  $0.0521^{+0.0079}_{-0.0083}$\\
$\Ntot$         &  $3.046$                        &  $2.96^{+0.27}_{-0.3}$          &  $3.197^{+0.039}_{-0.151}$      &  $2.96^{+0.28}_{-0.27}$\\
$\ffs$          &  $1$                            &  $1$                            &  $>0.88$      &  $>0.72$\\
\hline
$H_0$ [km/s/Mpc]               &  $67.69^{+0.73}_{-0.77}$    &  $67.1^{+2.1}_{-2.3}$           &  $68.87^{+0.87}_{-1.5}$         &  $67.2^{+2.1}_{-2.2}$\\
$r_s^\mathrm{drag}$ [Mpc]         &  $147.37^{+0.41}_{-0.35}$  &  $148.2^{+2.6}_{-2.8}$  &  $145.8^{+1.7}_{-0.7}$          &  $147.9^{+2.4}_{-2.7}$\\
$\sigma_8$          &  $0.8222^{+0.0064}_{-0.0063}$  &  $0.819^{+0.013}_{-0.013}$   &  $0.8252^{+0.0073}_{-0.0072}$   &  $0.813^{+0.013}_{-0.013}$\\
\hline
$\chi^2_{\mathrm{tot}}$ & 1189.2  &  1188.93  &  1189.11  &  1188.41 \\
\hline
	\end{tabular} 
	\caption{Comparison of extensions of $\Lambda$CDM with extra radiation degrees of freedom for the {\it Planck} TT likelihood.
        We present the marginalized mean $\pm 1 \, \sigma$ error for all the parameters, with the exception of $\ffs$ for which we show the $95\% $ lower limit. 
      The $\chi^2_{\mathrm{tot}}$ values correspond to best-fit point found using numerical minimization. 
	\label{tab:ntot_planck_TT_only}}
\end{table}

\section{Posterior Distributions}
\label{sec:posterior_distributions}

\begin{figure}
	\centering
	\includegraphics[width=0.95\textwidth]{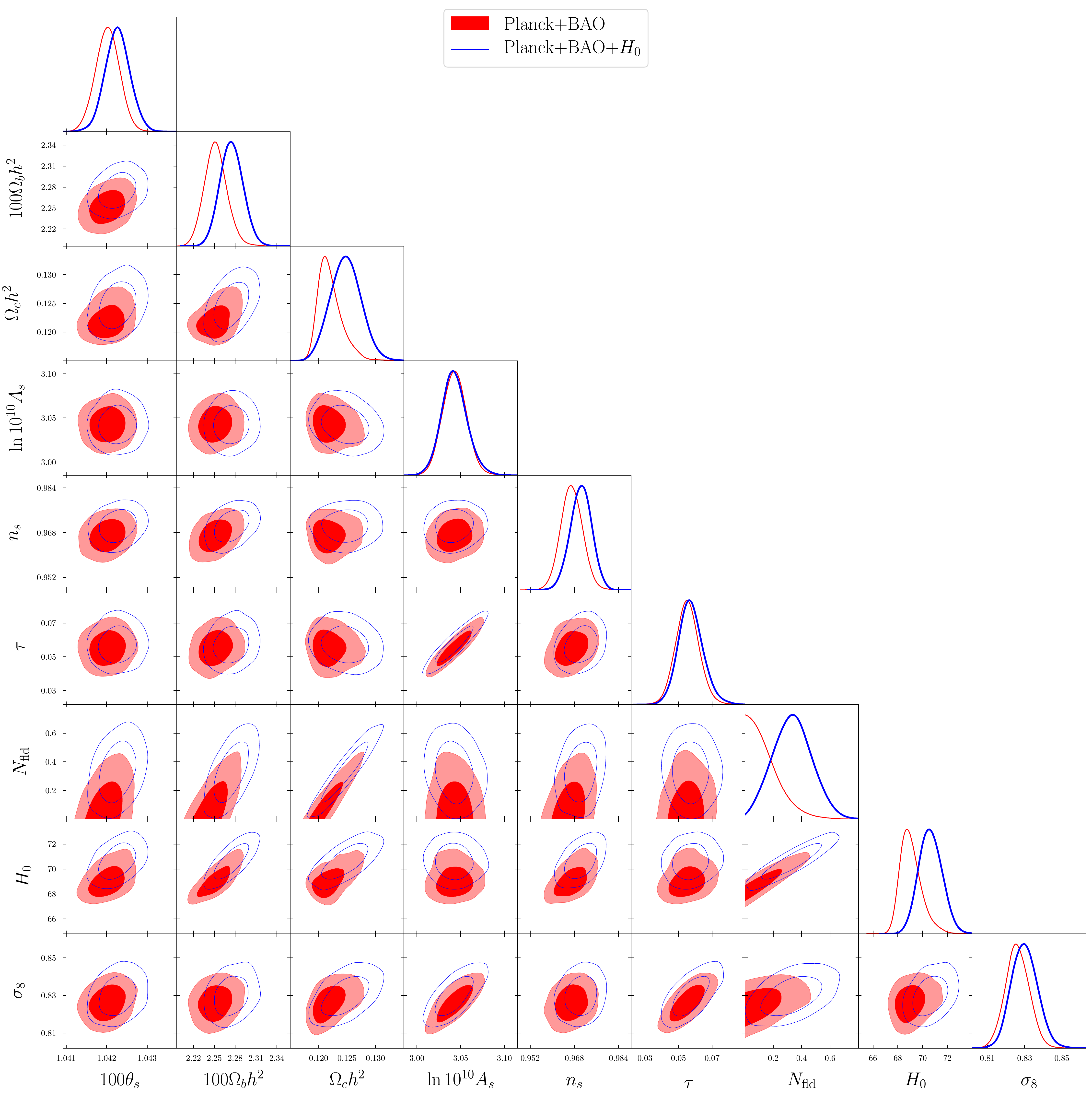}
    \caption{Posterior distributions for a model with $\Neff=3.046$ and varying $\Nfld$ for the {\it Planck} TT, TE, EE + BAO data set with (thin blue contours) and 
    without an $H_0$ prior (filled red regions). In both cases, the inner (outer) contours correspond to 68\% (95\%) confidence regions.}
	\label{fig:corner-nfld}
\end{figure}

\begin{figure}
	\centering
	\includegraphics[width=0.95\textwidth]{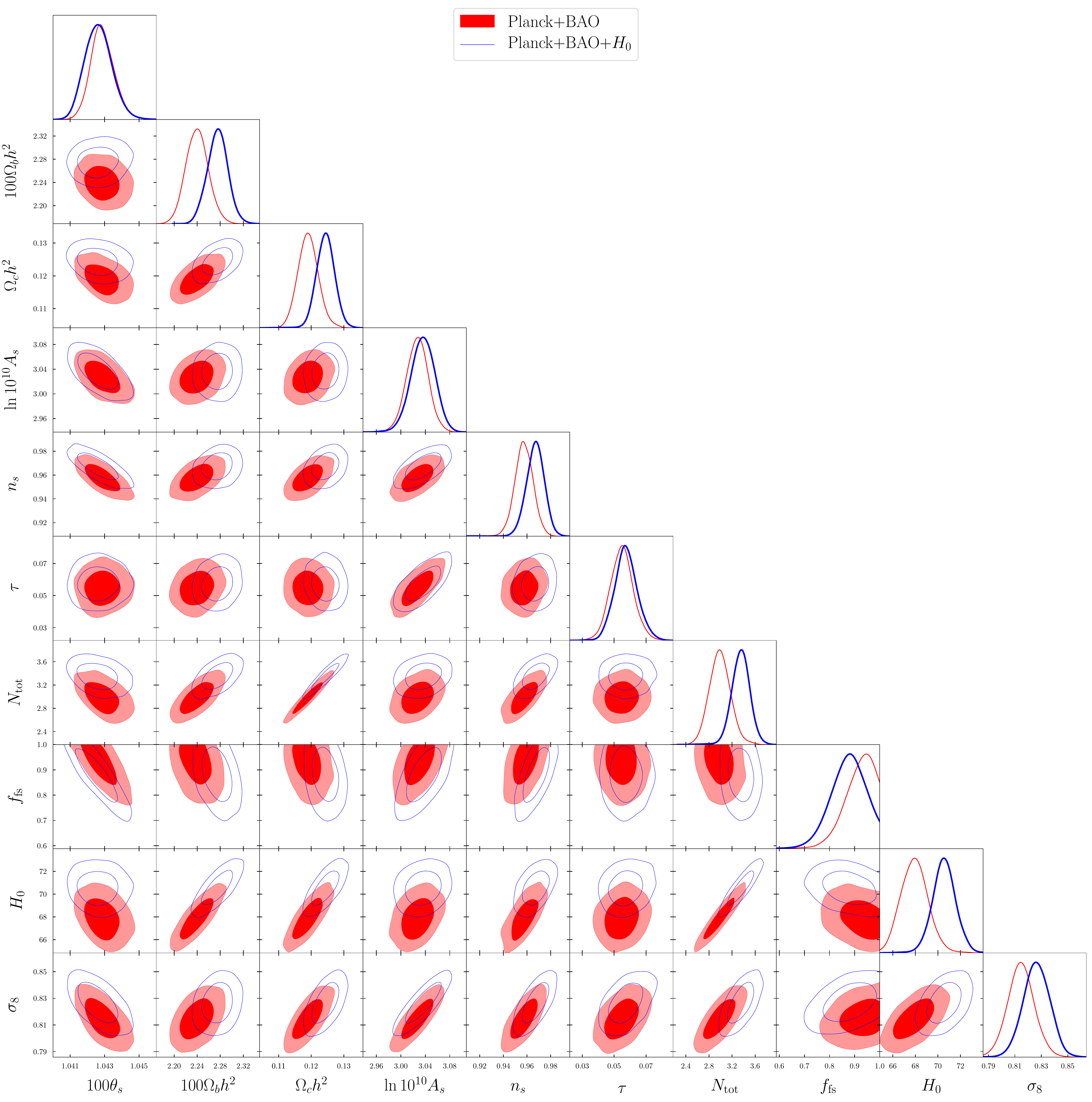}
	\caption{Posterior distributions for a model wit varying $\Ntot$ and $\ffs$ for the {\it Planck} TT, TE, EE + BAO data set with (thin blue contours) and 
    without an $H_0$ prior (filled red regions). In both cases, the inner (outer) contours correspond to 68\% (95\%) confidence regions.}
	\label{fig:corner-ntot}
\end{figure}

Here we present the full combination of 2d posterior distributions for the model with $\Nfld$ only (Fig.~\ref{fig:corner-nfld}) and with $\Neff$ + $\Nfld$ (Fig.~\ref{fig:corner-ntot}) using the {\it Planck} TT, TE, EE + BAO likelihood combination, and demonstrate the impact of an additional prior on $H_0$ from the 
local measurement of Ref.~\cite{Riess:2019cxk}.

\bibliographystyle{JHEP}
\bibliography{biblio}

\providecommand{\href}[2]{#2}\begingroup\raggedright\begin{thebibliography}{10}

\bibitem{Aghanim:2018eyx}
{\scshape Planck} collaboration, \emph{{Planck 2018 results. VI. Cosmological
  parameters}},  \href{https://arxiv.org/abs/1807.06209}{{\ttfamily
  1807.06209}}.

\bibitem{Abazajian:2001nj}
K.~Abazajian, G.~M. Fuller and M.~Patel, \emph{{Sterile neutrino hot, warm, and
  cold dark matter}},
  \href{https://doi.org/10.1103/PhysRevD.64.023501}{\emph{Phys. Rev.}
  {\bfseries D64} (2001) 023501}
  [\href{https://arxiv.org/abs/astro-ph/0101524}{{\ttfamily
  astro-ph/0101524}}].

\bibitem{Vogel:2013raa}
H.~Vogel and J.~Redondo, \emph{{Dark Radiation constraints on minicharged
  particles in models with a hidden photon}},
  \href{https://doi.org/10.1088/1475-7516/2014/02/029}{\emph{JCAP} {\bfseries
  1402} (2014) 029} [\href{https://arxiv.org/abs/1311.2600}{{\ttfamily
  1311.2600}}].

\bibitem{Berlin:2017ftj}
A.~Berlin and N.~Blinov, \emph{{Thermal Dark Matter Below an MeV}},
  \href{https://doi.org/10.1103/PhysRevLett.120.021801}{\emph{Phys. Rev. Lett.}
  {\bfseries 120} (2018) 021801}
  [\href{https://arxiv.org/abs/1706.07046}{{\ttfamily 1706.07046}}].

\bibitem{Halverson:2018xge}
J.~Halverson and P.~Langacker, \emph{{TASI Lectures on Remnants from the String
  Landscape}}, \href{https://doi.org/10.22323/1.305.0019}{\emph{PoS} {\bfseries
  TASI2017} (2018) 019} [\href{https://arxiv.org/abs/1801.03503}{{\ttfamily
  1801.03503}}].

\bibitem{Baumann:2016wac}
D.~Baumann, D.~Green and B.~Wallisch, \emph{{New Target for Cosmic Axion
  Searches}}, \href{https://doi.org/10.1103/PhysRevLett.117.171301}{\emph{Phys.
  Rev. Lett.} {\bfseries 117} (2016) 171301}
  [\href{https://arxiv.org/abs/1604.08614}{{\ttfamily 1604.08614}}].

\bibitem{Weinberg:2013kea}
S.~Weinberg, \emph{{Goldstone Bosons as Fractional Cosmic Neutrinos}},
  \href{https://doi.org/10.1103/PhysRevLett.110.241301}{\emph{Phys. Rev. Lett.}
  {\bfseries 110} (2013) 241301}
  [\href{https://arxiv.org/abs/1305.1971}{{\ttfamily 1305.1971}}].

\bibitem{Millea:2015qra}
M.~Millea, L.~Knox and B.~Fields, \emph{{New Bounds for Axions and Axion-Like
  Particles with keV-GeV Masses}},
  \href{https://doi.org/10.1103/PhysRevD.92.023010}{\emph{Phys. Rev.}
  {\bfseries D92} (2015) 023010}
  [\href{https://arxiv.org/abs/1501.04097}{{\ttfamily 1501.04097}}].

\bibitem{Abazajian:2016yjj}
{\scshape CMB-S4} collaboration, \emph{{CMB-S4 Science Book, First Edition}},
  \href{https://arxiv.org/abs/1610.02743}{{\ttfamily 1610.02743}}.

\bibitem{Ade:2018sbj}
{\scshape Simons Observatory} collaboration, \emph{{The Simons Observatory:
  Science goals and forecasts}},
  \href{https://doi.org/10.1088/1475-7516/2019/02/056}{\emph{JCAP} {\bfseries
  1902} (2019) 056} [\href{https://arxiv.org/abs/1808.07445}{{\ttfamily
  1808.07445}}].

\bibitem{Bashinsky:2003tk}
S.~Bashinsky and U.~Seljak, \emph{{Neutrino perturbations in CMB anisotropy and
  matter clustering}},
  \href{https://doi.org/10.1103/PhysRevD.69.083002}{\emph{Phys. Rev.}
  {\bfseries D69} (2004) 083002}
  [\href{https://arxiv.org/abs/astro-ph/0310198}{{\ttfamily
  astro-ph/0310198}}].

\bibitem{Hou:2011ec}
Z.~Hou, R.~Keisler, L.~Knox, M.~Millea and C.~Reichardt, \emph{{How Massless
  Neutrinos Affect the Cosmic Microwave Background Damping Tail}},
  \href{https://doi.org/10.1103/PhysRevD.87.083008}{\emph{Phys. Rev.}
  {\bfseries D87} (2013) 083008}
  [\href{https://arxiv.org/abs/1104.2333}{{\ttfamily 1104.2333}}].

\bibitem{Baumann:2015rya}
D.~Baumann, D.~Green, J.~Meyers and B.~Wallisch, \emph{{Phases of New Physics
  in the CMB}},
  \href{https://doi.org/10.1088/1475-7516/2016/01/007}{\emph{JCAP} {\bfseries
  1601} (2016) 007} [\href{https://arxiv.org/abs/1508.06342}{{\ttfamily
  1508.06342}}].

\bibitem{Friedland:2007vv}
A.~Friedland, K.~M. Zurek and S.~Bashinsky, \emph{{Constraining Models of
  Neutrino Mass and Neutrino Interactions with the Planck Satellite}},
  \href{https://arxiv.org/abs/0704.3271}{{\ttfamily 0704.3271}}.

\bibitem{Follin:2015hya}
B.~Follin, L.~Knox, M.~Millea and Z.~Pan, \emph{{First Detection of the
  Acoustic Oscillation Phase Shift Expected from the Cosmic Neutrino
  Background}},
  \href{https://doi.org/10.1103/PhysRevLett.115.091301}{\emph{Phys. Rev. Lett.}
  {\bfseries 115} (2015) 091301}
  [\href{https://arxiv.org/abs/1503.07863}{{\ttfamily 1503.07863}}].

\bibitem{Riess:2019cxk}
A.~G. Riess, S.~Casertano, W.~Yuan, L.~M. Macri and D.~Scolnic, \emph{{Large
  Magellanic Cloud Cepheid Standards Provide a 1\% Foundation for the
  Determination of the Hubble Constant and Stronger Evidence for Physics beyond
  $\Lambda$CDM}},
  \href{https://doi.org/10.3847/1538-4357/ab1422}{\emph{Astrophys. J.}
  {\bfseries 876} (2019) 85}
  [\href{https://arxiv.org/abs/1903.07603}{{\ttfamily 1903.07603}}].

\bibitem{Wong:2019kwg}
K.~C. Wong et~al., \emph{{H0LiCOW XIII. A 2.4\% measurement of $H_{0}$ from
  lensed quasars: $5.3\sigma$ tension between early and late-Universe probes}},
   \href{https://arxiv.org/abs/1907.04869}{{\ttfamily 1907.04869}}.

\bibitem{Abbott:2017smn}
{\scshape DES} collaboration, \emph{{Dark Energy Survey Year 1 Results: A
  Precise H0 Estimate from DES Y1, BAO, and D/H Data}},
  \href{https://doi.org/10.1093/mnras/sty1939}{\emph{Mon. Not. Roy. Astron.
  Soc.} {\bfseries 480} (2018) 3879}
  [\href{https://arxiv.org/abs/1711.00403}{{\ttfamily 1711.00403}}].

\bibitem{Addison:2017fdm}
G.~E. Addison, D.~J. Watts, C.~L. Bennett, M.~Halpern, G.~Hinshaw and J.~L.
  Weiland, \emph{{Elucidating $\Lambda$CDM: Impact of Baryon Acoustic
  Oscillation Measurements on the Hubble Constant Discrepancy}},
  \href{https://doi.org/10.3847/1538-4357/aaa1ed}{\emph{Astrophys. J.}
  {\bfseries 853} (2018) 119}
  [\href{https://arxiv.org/abs/1707.06547}{{\ttfamily 1707.06547}}].

\bibitem{Schoneberg:2019wmt}
N.~Schöneberg, J.~Lesgourgues and D.~C. Hooper, \emph{{The BAO+BBN take on the
  Hubble tension}},
  \href{https://doi.org/10.1088/1475-7516/2019/10/029}{\emph{JCAP} {\bfseries
  1910} (2019) 029} [\href{https://arxiv.org/abs/1907.11594}{{\ttfamily
  1907.11594}}].

\bibitem{Freedman:2019jwv}
W.~L. Freedman et~al., \emph{{The Carnegie-Chicago Hubble Program. VIII. An
  Independent Determination of the Hubble Constant Based on the Tip of the Red
  Giant Branch}},  \href{https://arxiv.org/abs/1907.05922}{{\ttfamily
  1907.05922}}.

\bibitem{Bernal:2016gxb}
J.~L. Bernal, L.~Verde and A.~G. Riess, \emph{{The trouble with $H_0$}},
  \href{https://doi.org/10.1088/1475-7516/2016/10/019}{\emph{JCAP} {\bfseries
  1610} (2016) 019} [\href{https://arxiv.org/abs/1607.05617}{{\ttfamily
  1607.05617}}].

\bibitem{Aylor:2018drw}
K.~Aylor, M.~Joy, L.~Knox, M.~Millea, S.~Raghunathan and W.~L.~K. Wu,
  \emph{{Sounds Discordant: Classical Distance Ladder \& $\Lambda$CDM -based
  Determinations of the Cosmological Sound Horizon}},
  \href{https://doi.org/10.3847/1538-4357/ab0898}{\emph{Astrophys. J.}
  {\bfseries 874} (2019) 4} [\href{https://arxiv.org/abs/1811.00537}{{\ttfamily
  1811.00537}}].

\bibitem{Knox:2019rjx}
L.~Knox and M.~Millea, \emph{{Hubble constant hunter’s guide}},
  \href{https://doi.org/10.1103/PhysRevD.101.043533}{\emph{Phys. Rev.}
  {\bfseries D101} (2020) 043533}
  [\href{https://arxiv.org/abs/1908.03663}{{\ttfamily 1908.03663}}].

\bibitem{Riess:2016jrr}
A.~G. Riess et~al., \emph{{A 2.4\% Determination of the Local Value of the
  Hubble Constant}},
  \href{https://doi.org/10.3847/0004-637X/826/1/56}{\emph{Astrophys. J.}
  {\bfseries 826} (2016) 56}
  [\href{https://arxiv.org/abs/1604.01424}{{\ttfamily 1604.01424}}].

\bibitem{Kreisch:2019yzn}
C.~D. Kreisch, F.-Y. Cyr-Racine and O.~Doré, \emph{{The Neutrino Puzzle:
  Anomalies, Interactions, and Cosmological Tensions}},
  \href{https://arxiv.org/abs/1902.00534}{{\ttfamily 1902.00534}}.

\bibitem{Forastieri:2019cuf}
F.~Forastieri, M.~Lattanzi and P.~Natoli, \emph{{Cosmological constraints on
  neutrino self-interactions with a light mediator}},
  \href{https://doi.org/10.1103/PhysRevD.100.103526}{\emph{Phys. Rev.}
  {\bfseries D100} (2019) 103526}
  [\href{https://arxiv.org/abs/1904.07810}{{\ttfamily 1904.07810}}].

\bibitem{DiValentino:2017oaw}
E.~Di~Valentino, C.~Bøehm, E.~Hivon and F.~R. Bouchet, \emph{{Reducing the
  $H_0$ and $\sigma_8$ tensions with Dark Matter-neutrino interactions}},
  \href{https://doi.org/10.1103/PhysRevD.97.043513}{\emph{Phys. Rev.}
  {\bfseries D97} (2018) 043513}
  [\href{https://arxiv.org/abs/1710.02559}{{\ttfamily 1710.02559}}].

\bibitem{Ghosh:2019tab}
S.~Ghosh, R.~Khatri and T.~S. Roy, \emph{{Dark Neutrino interactions phase out
  the Hubble tension}},  \href{https://arxiv.org/abs/1908.09843}{{\ttfamily
  1908.09843}}.

\bibitem{Poulin:2016nat}
V.~Poulin, P.~D. Serpico and J.~Lesgourgues, \emph{{A fresh look at linear
  cosmological constraints on a decaying dark matter component}},
  \href{https://doi.org/10.1088/1475-7516/2016/08/036}{\emph{JCAP} {\bfseries
  1608} (2016) 036} [\href{https://arxiv.org/abs/1606.02073}{{\ttfamily
  1606.02073}}].

\bibitem{Poulin:2018cxd}
V.~Poulin, T.~L. Smith, T.~Karwal and M.~Kamionkowski, \emph{{Early Dark Energy
  Can Resolve The Hubble Tension}},
  \href{https://arxiv.org/abs/1811.04083}{{\ttfamily 1811.04083}}.

\bibitem{Poulin:2018dzj}
V.~Poulin, T.~L. Smith, D.~Grin, T.~Karwal and M.~Kamionkowski,
  \emph{{Cosmological implications of ultralight axionlike fields}},
  \href{https://doi.org/10.1103/PhysRevD.98.083525}{\emph{Phys. Rev.}
  {\bfseries D98} (2018) 083525}
  [\href{https://arxiv.org/abs/1806.10608}{{\ttfamily 1806.10608}}].

\bibitem{Agrawal:2019lmo}
P.~Agrawal, F.-Y. Cyr-Racine, D.~Pinner and L.~Randall, \emph{{Rock 'n' Roll
  Solutions to the Hubble Tension}},
  \href{https://arxiv.org/abs/1904.01016}{{\ttfamily 1904.01016}}.

\bibitem{Smith:2019ihp}
T.~L. Smith, V.~Poulin and M.~A. Amin, \emph{{Oscillating scalar fields and the
  Hubble tension: a resolution with novel signatures}},
  \href{https://arxiv.org/abs/1908.06995}{{\ttfamily 1908.06995}}.

\bibitem{Abbott:2017wau}
{\scshape DES} collaboration, \emph{{Dark Energy Survey year 1 results:
  Cosmological constraints from galaxy clustering and weak lensing}},
  \href{https://doi.org/10.1103/PhysRevD.98.043526}{\emph{Phys. Rev.}
  {\bfseries D98} (2018) 043526}
  [\href{https://arxiv.org/abs/1708.01530}{{\ttfamily 1708.01530}}].

\bibitem{Hildebrandt:2018yau}
H.~Hildebrandt et~al., \emph{{KiDS+VIKING-450: Cosmic shear tomography with
  optical+infrared data}},
  \href{https://doi.org/10.1051/0004-6361/201834878}{\emph{Astron. Astrophys.}
  {\bfseries 633} (2020) A69}
  [\href{https://arxiv.org/abs/1812.06076}{{\ttfamily 1812.06076}}].

\bibitem{Abbott:2018wzc}
{\scshape DES} collaboration, \emph{{Cosmological Constraints from Multiple
  Probes in the Dark Energy Survey}},
  \href{https://doi.org/10.1103/PhysRevLett.122.171301}{\emph{Phys. Rev. Lett.}
  {\bfseries 122} (2019) 171301}
  [\href{https://arxiv.org/abs/1811.02375}{{\ttfamily 1811.02375}}].

\bibitem{Hikage:2018qbn}
{\scshape HSC} collaboration, \emph{{Cosmology from cosmic shear power spectra
  with Subaru Hyper Suprime-Cam first-year data}},
  \href{https://doi.org/10.1093/pasj/psz010}{\emph{Publ. Astron. Soc. Jap.}
  {\bfseries 71} (2019) Publications of the Astronomical Society of Japan,
  Volume 71, Issue 2, April 2019, 43, https://doi.org/10.1093/pasj/psz010}
  [\href{https://arxiv.org/abs/1809.09148}{{\ttfamily 1809.09148}}].

\bibitem{Blas:2011rf}
D.~Blas, J.~Lesgourgues and T.~Tram, \emph{{The Cosmic Linear Anisotropy
  Solving System (CLASS) II: Approximation schemes}},
  \href{https://doi.org/10.1088/1475-7516/2011/07/034}{\emph{JCAP} {\bfseries
  1107} (2011) 034} [\href{https://arxiv.org/abs/1104.2933}{{\ttfamily
  1104.2933}}].

\bibitem{Lewis:1999bs}
A.~Lewis, A.~Challinor and A.~Lasenby, \emph{{Efficient computation of CMB
  anisotropies in closed FRW models}},
  \href{https://doi.org/10.1086/309179}{\emph{Astrophys. J.} {\bfseries 538}
  (2000) 473} [\href{https://arxiv.org/abs/astro-ph/9911177}{{\ttfamily
  astro-ph/9911177}}].

\bibitem{Brust:2017nmv}
C.~Brust, Y.~Cui and K.~Sigurdson, \emph{{Cosmological Constraints on
  Interacting Light Particles}},
  \href{https://doi.org/10.1088/1475-7516/2017/08/020}{\emph{JCAP} {\bfseries
  1708} (2017) 020} [\href{https://arxiv.org/abs/1703.10732}{{\ttfamily
  1703.10732}}].

\bibitem{Aghanim:2019ame}
{\scshape Planck} collaboration, \emph{{Planck 2018 results. V. CMB power
  spectra and likelihoods}},
  \href{https://arxiv.org/abs/1907.12875}{{\ttfamily 1907.12875}}.

\bibitem{Escudero:2019gvw}
M.~Escudero and S.~J. Witte, \emph{{A CMB Search for the Neutrino Mass
  Mechanism and its Relation to the $H_0$ Tension}},
  \href{https://arxiv.org/abs/1909.04044}{{\ttfamily 1909.04044}}.

\bibitem{Mangano:2005cc}
G.~Mangano, G.~Miele, S.~Pastor, T.~Pinto, O.~Pisanti and P.~D. Serpico,
  \emph{{Relic neutrino decoupling including flavor oscillations}},
  \href{https://doi.org/10.1016/j.nuclphysb.2005.09.041}{\emph{Nucl. Phys.}
  {\bfseries B729} (2005) 221}
  [\href{https://arxiv.org/abs/hep-ph/0506164}{{\ttfamily hep-ph/0506164}}].

\bibitem{Grohs:2015tfy}
E.~Grohs, G.~M. Fuller, C.~T. Kishimoto, M.~W. Paris and A.~Vlasenko,
  \emph{{Neutrino energy transport in weak decoupling and big bang
  nucleosynthesis}},
  \href{https://doi.org/10.1103/PhysRevD.93.083522}{\emph{Phys. Rev.}
  {\bfseries D93} (2016) 083522}
  [\href{https://arxiv.org/abs/1512.02205}{{\ttfamily 1512.02205}}].

\bibitem{deSalas:2016ztq}
P.~F. de~Salas and S.~Pastor, \emph{{Relic neutrino decoupling with flavour
  oscillations revisited}},
  \href{https://doi.org/10.1088/1475-7516/2016/07/051}{\emph{JCAP} {\bfseries
  1607} (2016) 051} [\href{https://arxiv.org/abs/1606.06986}{{\ttfamily
  1606.06986}}].

\bibitem{Jeong:2013eza}
K.~S. Jeong and F.~Takahashi, \emph{{Self-interacting Dark Radiation}},
  \href{https://doi.org/10.1016/j.physletb.2013.07.001}{\emph{Phys. Lett.}
  {\bfseries B725} (2013) 134}
  [\href{https://arxiv.org/abs/1305.6521}{{\ttfamily 1305.6521}}].

\bibitem{Buen-Abad:2015ova}
M.~A. Buen-Abad, G.~Marques-Tavares and M.~Schmaltz, \emph{{Non-Abelian dark
  matter and dark radiation}},
  \href{https://doi.org/10.1103/PhysRevD.92.023531}{\emph{Phys. Rev.}
  {\bfseries D92} (2015) 023531}
  [\href{https://arxiv.org/abs/1505.03542}{{\ttfamily 1505.03542}}].

\bibitem{Lesgourgues:2015wza}
J.~Lesgourgues, G.~Marques-Tavares and M.~Schmaltz, \emph{{Evidence for dark
  matter interactions in cosmological precision data?}},
  \href{https://doi.org/10.1088/1475-7516/2016/02/037}{\emph{JCAP} {\bfseries
  1602} (2016) 037} [\href{https://arxiv.org/abs/1507.04351}{{\ttfamily
  1507.04351}}].

\bibitem{Buen-Abad:2017gxg}
M.~A. Buen-Abad, M.~Schmaltz, J.~Lesgourgues and T.~Brinckmann,
  \emph{{Interacting Dark Sector and Precision Cosmology}},
  \href{https://doi.org/10.1088/1475-7516/2018/01/008}{\emph{JCAP} {\bfseries
  1801} (2018) 008} [\href{https://arxiv.org/abs/1708.09406}{{\ttfamily
  1708.09406}}].

\bibitem{Chacko:2003dt}
Z.~Chacko, L.~J. Hall, T.~Okui and S.~J. Oliver, \emph{{CMB signals of neutrino
  mass generation}},
  \href{https://doi.org/10.1103/PhysRevD.70.085008}{\emph{Phys. Rev.}
  {\bfseries D70} (2004) 085008}
  [\href{https://arxiv.org/abs/hep-ph/0312267}{{\ttfamily hep-ph/0312267}}].

\bibitem{Chacko:2004cz}
Z.~Chacko, L.~J. Hall, S.~J. Oliver and M.~Perelstein, \emph{{Late time
  neutrino masses, the LSND experiment and the cosmic microwave background}},
  \href{https://doi.org/10.1103/PhysRevLett.94.111801}{\emph{Phys. Rev. Lett.}
  {\bfseries 94} (2005) 111801}
  [\href{https://arxiv.org/abs/hep-ph/0405067}{{\ttfamily hep-ph/0405067}}].

\bibitem{Berlin:2018ztp}
A.~Berlin and N.~Blinov, \emph{{Thermal neutrino portal to sub-MeV dark
  matter}}, \href{https://doi.org/10.1103/PhysRevD.99.095030}{\emph{Phys. Rev.}
  {\bfseries D99} (2019) 095030}
  [\href{https://arxiv.org/abs/1807.04282}{{\ttfamily 1807.04282}}].

\bibitem{Berlin:2019pbq}
A.~Berlin, N.~Blinov and S.~W. Li, \emph{{Dark Sector Equilibration During
  Nucleosynthesis}},
  \href{https://doi.org/10.1103/PhysRevD.100.015038}{\emph{Phys. Rev.}
  {\bfseries D100} (2019) 015038}
  [\href{https://arxiv.org/abs/1904.04256}{{\ttfamily 1904.04256}}].

\bibitem{Hannestad:2005ex}
S.~Hannestad and G.~Raffelt, \emph{{Constraining invisible neutrino decays with
  the cosmic microwave background}},
  \href{https://doi.org/10.1103/PhysRevD.72.103514}{\emph{Phys. Rev.}
  {\bfseries D72} (2005) 103514}
  [\href{https://arxiv.org/abs/hep-ph/0509278}{{\ttfamily hep-ph/0509278}}].

\bibitem{Cyr-Racine:2013jua}
F.-Y. Cyr-Racine and K.~Sigurdson, \emph{{Limits on Neutrino-Neutrino
  Scattering in the Early Universe}},
  \href{https://doi.org/10.1103/PhysRevD.90.123533}{\emph{Phys. Rev.}
  {\bfseries D90} (2014) 123533}
  [\href{https://arxiv.org/abs/1306.1536}{{\ttfamily 1306.1536}}].

\bibitem{Archidiacono:2013dua}
M.~Archidiacono and S.~Hannestad, \emph{{Updated constraints on non-standard
  neutrino interactions from Planck}},
  \href{https://doi.org/10.1088/1475-7516/2014/07/046}{\emph{JCAP} {\bfseries
  07} (2014) 046} [\href{https://arxiv.org/abs/1311.3873}{{\ttfamily
  1311.3873}}].

\bibitem{Oldengott:2014qra}
I.~M. Oldengott, C.~Rampf and Y.~Y.~Y. Wong, \emph{{Boltzmann hierarchy for
  interacting neutrinos I: formalism}},
  \href{https://doi.org/10.1088/1475-7516/2015/04/016}{\emph{JCAP} {\bfseries
  1504} (2015) 016} [\href{https://arxiv.org/abs/1409.1577}{{\ttfamily
  1409.1577}}].

\bibitem{Forastieri:2015paa}
F.~Forastieri, M.~Lattanzi and P.~Natoli, \emph{{Constraints on secret neutrino
  interactions after Planck}},
  \href{https://doi.org/10.1088/1475-7516/2015/07/014}{\emph{JCAP} {\bfseries
  1507} (2015) 014} [\href{https://arxiv.org/abs/1504.04999}{{\ttfamily
  1504.04999}}].

\bibitem{Oldengott:2017fhy}
I.~M. Oldengott, T.~Tram, C.~Rampf and Y.~Y.~Y. Wong, \emph{{Interacting
  neutrinos in cosmology: exact description and constraints}},
  \href{https://doi.org/10.1088/1475-7516/2017/11/027}{\emph{JCAP} {\bfseries
  1711} (2017) 027} [\href{https://arxiv.org/abs/1706.02123}{{\ttfamily
  1706.02123}}].

\bibitem{Lancaster:2017ksf}
L.~Lancaster, F.-Y. Cyr-Racine, L.~Knox and Z.~Pan, \emph{{A tale of two modes:
  Neutrino free-streaming in the early universe}},
  \href{https://doi.org/10.1088/1475-7516/2017/07/033}{\emph{JCAP} {\bfseries
  1707} (2017) 033} [\href{https://arxiv.org/abs/1704.06657}{{\ttfamily
  1704.06657}}].

\bibitem{Blinov:2019gcj}
N.~Blinov, K.~J. Kelly, G.~Z. Krnjaic and S.~D. McDermott, \emph{{Constraining
  the Self-Interacting Neutrino Interpretation of the Hubble Tension}},
  \href{https://doi.org/10.1103/PhysRevLett.123.191102}{\emph{Phys. Rev. Lett.}
  {\bfseries 123} (2019) 191102}
  [\href{https://arxiv.org/abs/1905.02727}{{\ttfamily 1905.02727}}].

\bibitem{Beacom:2004yd}
J.~F. Beacom, N.~F. Bell and S.~Dodelson, \emph{{Neutrinoless universe}},
  \href{https://doi.org/10.1103/PhysRevLett.93.121302}{\emph{Phys. Rev. Lett.}
  {\bfseries 93} (2004) 121302}
  [\href{https://arxiv.org/abs/astro-ph/0404585}{{\ttfamily
  astro-ph/0404585}}].

\bibitem{Choi:2018gho}
G.~Choi, C.-T. Chiang and M.~LoVerde, \emph{{Probing Decoupling in Dark Sectors
  with the Cosmic Microwave Background}},
  \href{https://doi.org/10.1088/1475-7516/2018/06/044}{\emph{JCAP} {\bfseries
  1806} (2018) 044} [\href{https://arxiv.org/abs/1804.10180}{{\ttfamily
  1804.10180}}].

\bibitem{Ma:1995ey}
C.-P. Ma and E.~Bertschinger, \emph{{Cosmological perturbation theory in the
  synchronous and conformal Newtonian gauges}},
  \href{https://doi.org/10.1086/176550}{\emph{Astrophys. J.} {\bfseries 455}
  (1995) 7} [\href{https://arxiv.org/abs/astro-ph/9506072}{{\ttfamily
  astro-ph/9506072}}].

\bibitem{Hu:2004xd}
W.~Hu, \emph{{Covariant linear perturbation formalism}}, {\emph{ICTP Lect.
  Notes Ser.} {\bfseries 14} (2003) 145}
  [\href{https://arxiv.org/abs/astro-ph/0402060}{{\ttfamily
  astro-ph/0402060}}].

\bibitem{Bond:1983hb}
J.~R. Bond and A.~S. Szalay, \emph{{The Collisionless Damping of Density
  Fluctuations in an Expanding Universe}},
  \href{https://doi.org/10.1086/161460}{\emph{Astrophys. J.} {\bfseries 274}
  (1983) 443}.

\bibitem{Hu:1994uz}
W.~Hu and N.~Sugiyama, \emph{{Anisotropies in the cosmic microwave background:
  An Analytic approach}},
  \href{https://doi.org/10.1086/175624}{\emph{Astrophys. J.} {\bfseries 444}
  (1995) 489} [\href{https://arxiv.org/abs/astro-ph/9407093}{{\ttfamily
  astro-ph/9407093}}].

\bibitem{Zaldarriaga:1995gi}
M.~Zaldarriaga and D.~D. Harari, \emph{{Analytic approach to the polarization
  of the cosmic microwave background in flat and open universes}},
  \href{https://doi.org/10.1103/PhysRevD.52.3276}{\emph{Phys. Rev.} {\bfseries
  D52} (1995) 3276} [\href{https://arxiv.org/abs/astro-ph/9504085}{{\ttfamily
  astro-ph/9504085}}].

\bibitem{Hu:1996mn}
W.~Hu and M.~J. White, \emph{{The Damping tail of CMB anisotropies}},
  \href{https://doi.org/10.1086/303928}{\emph{Astrophys. J.} {\bfseries 479}
  (1997) 568} [\href{https://arxiv.org/abs/astro-ph/9609079}{{\ttfamily
  astro-ph/9609079}}].

\bibitem{Hu:2000ti}
W.~Hu, M.~Fukugita, M.~Zaldarriaga and M.~Tegmark, \emph{{CMB observables and
  their cosmological implications}},
  \href{https://doi.org/10.1086/319449}{\emph{Astrophys. J.} {\bfseries 549}
  (2001) 669} [\href{https://arxiv.org/abs/astro-ph/0006436}{{\ttfamily
  astro-ph/0006436}}].

\bibitem{Hu:1995en}
W.~Hu and N.~Sugiyama, \emph{{Small scale cosmological perturbations: An
  Analytic approach}}, \href{https://doi.org/10.1086/177989}{\emph{Astrophys.
  J.} {\bfseries 471} (1996) 542}
  [\href{https://arxiv.org/abs/astro-ph/9510117}{{\ttfamily
  astro-ph/9510117}}].

\bibitem{Brinckmann:2018cvx}
T.~Brinckmann and J.~Lesgourgues, \emph{{MontePython 3: boosted MCMC sampler
  and other features}},  \href{https://arxiv.org/abs/1804.07261}{{\ttfamily
  1804.07261}}.

\bibitem{Alam:2016hwk}
{\scshape BOSS} collaboration, \emph{{The clustering of galaxies in the
  completed SDSS-III Baryon Oscillation Spectroscopic Survey: cosmological
  analysis of the DR12 galaxy sample}},
  \href{https://doi.org/10.1093/mnras/stx721}{\emph{Mon. Not. Roy. Astron.
  Soc.} {\bfseries 470} (2017) 2617}
  [\href{https://arxiv.org/abs/1607.03155}{{\ttfamily 1607.03155}}].

\bibitem{Beutler:2011hx}
F.~Beutler, C.~Blake, M.~Colless, D.~H. Jones, L.~Staveley-Smith, L.~Campbell
  et~al., \emph{{The 6dF Galaxy Survey: Baryon Acoustic Oscillations and the
  Local Hubble Constant}},
  \href{https://doi.org/10.1111/j.1365-2966.2011.19250.x}{\emph{Mon. Not. Roy.
  Astron. Soc.} {\bfseries 416} (2011) 3017}
  [\href{https://arxiv.org/abs/1106.3366}{{\ttfamily 1106.3366}}].

\bibitem{Ross:2014qpa}
A.~J. Ross, L.~Samushia, C.~Howlett, W.~J. Percival, A.~Burden and M.~Manera,
  \emph{{The clustering of the SDSS DR7 main Galaxy sample – I. A 4 per cent
  distance measure at $z = 0.15$}},
  \href{https://doi.org/10.1093/mnras/stv154}{\emph{Mon. Not. Roy. Astron.
  Soc.} {\bfseries 449} (2015) 835}
  [\href{https://arxiv.org/abs/1409.3242}{{\ttfamily 1409.3242}}].

\bibitem{Gelman:1992zz}
A.~Gelman and D.~B. Rubin, \emph{{Inference from Iterative Simulation Using
  Multiple Sequences}},
  \href{https://doi.org/10.1214/ss/1177011136}{\emph{Statist. Sci.} {\bfseries
  7} (1992) 457}.

\bibitem{Brooks:1998ab}
S.~P. Brooks and A.~Gelman, \emph{General methods for monitoring convergence of
  iterative simulations},
  \href{https://doi.org/10.1080/10618600.1998.10474787}{\emph{Journal of
  Computational and Graphical Statistics} {\bfseries 7} (1998) 434}
  [\href{https://arxiv.org/abs/https://amstat.tandfonline.com/doi/pdf/10.1080/10618600.1998.10474787}{{\ttfamily
  https://amstat.tandfonline.com/doi/pdf/10.1080/10618600.1998.10474787}}].

\bibitem{2020SciPy-NMeth}
P.~{Virtanen}, R.~{Gommers}, T.~E. {Oliphant}, M.~{Haberland}, T.~{Reddy},
  D.~{Cournapeau} et~al., \emph{{SciPy 1.0: Fundamental Algorithms for
  Scientific Computing in Python}},
  \href{https://doi.org/https://doi.org/10.1038/s41592-019-0686-2}{\emph{Nature
  Methods} {\bfseries 17} (2020) 261}.

\bibitem{iminuit}
iminuit team, ``iminuit -- a python interface to minuit.''
  \url{https://github.com/scikit-hep/iminuit}.

\bibitem{1975CoPhC..10..343J}
F.~{James} and M.~{Roos}, \emph{{Minuit -- a system for function minimization
  and analysis of the parameter errors and correlations}},
  \href{https://doi.org/10.1016/0010-4655(75)90039-9}{\emph{Computer Physics
  Communications} {\bfseries 10} (1975) 343}.

\bibitem{Ade:2015xua}
{\scshape Planck} collaboration, \emph{{Planck 2015 results. XIII. Cosmological
  parameters}},
  \href{https://doi.org/10.1051/0004-6361/201525830}{\emph{Astron. Astrophys.}
  {\bfseries 594} (2016) A13}
  [\href{https://arxiv.org/abs/1502.01589}{{\ttfamily 1502.01589}}].

\bibitem{Aver:2015iza}
E.~Aver, K.~A. Olive and E.~D. Skillman, \emph{{The effects of He I λ10830 on
  helium abundance determinations}},
  \href{https://doi.org/10.1088/1475-7516/2015/07/011}{\emph{JCAP} {\bfseries
  1507} (2015) 011} [\href{https://arxiv.org/abs/1503.08146}{{\ttfamily
  1503.08146}}].

\bibitem{Lewis:2019xzd}
A.~Lewis, \emph{{GetDist: a Python package for analysing Monte Carlo samples}},
   \href{https://arxiv.org/abs/1910.13970}{{\ttfamily 1910.13970}}.

\end{thebibliography}\endgroup
\end{document}